\documentclass[longauth, desactivate]{aa}
\usepackage{graphicx, array}
\usepackage[dvipsnames]{xcolor}
\usepackage[varg]{txfonts}
\usepackage{amssymb}
\usepackage{amsmath}
\usepackage[english]{babel}
\usepackage[style=british]{csquotes}         
\usepackage{nicefrac}
\usepackage{stfloats}
\usepackage{multirow}
\usepackage{booktabs}
\usepackage{placeins}
\usepackage{siunitx}
\usepackage[font=small]{caption}
\usepackage{subcaption}
\usepackage{ragged2e}       

\usepackage{hyperref}
\hypersetup{colorlinks = true,
            linkcolor = blue,
            urlcolor  = blue,
            citecolor = blue}

\makeatletter
\renewcommand*\aa@pageof{, page \thepage{} of \pageref*{LastPage}}
\makeatother

\newcommand{\Msun}{M_\odot}
\newcommand{\Md}{$M_\mathrm{d}$}
\newcommand{\Lsun}{L_\odot}

\newcommand{\Ha}{$H_\mathrm{\alpha}$}
\newcommand{\Qphi}{$ Q_\phi $}
\newcommand{\rout}{$r_\mathrm{out}$}
\newcommand{\rin}{$r_\mathrm{in}$}
\newcommand{\secname}{Sec.}
\newcommand{\eqname}{Eq.}
\newcolumntype{C}{>{$}c<{$}}    

\begin{document}

    \title{Disk Evolution Study Through Imaging of Nearby Young Stars (DESTINYS): V721 CrA and BN CrA have wide and structured disks in the polarised infrared}
    \author{
        G. Columba  \inst{1}
        \and
        E. Rigliaco \inst{2}
        \and
        R. Gratton \inst{2}
        \and
        C. Ginski \inst{3}
        \and
        A. Garufi  \inst{4}
        \and
        M. Benisty \inst{5,6}
        \and
        S. Facchini \inst{7}
        \and
        R.G. van Holstein \inst{8}
        \and
        A. Ribas  \inst{9}
        \and
        J. Williams  \inst{10}
        \and
        A. Zurlo  \inst{11,12}
        }
    
    \institute{ 
        Alma Mater Studiorum - University of Bologna, Dipartimento di Fisica e Astronomia “Augusto Righi”, Via Gobetti 93/2, 40129 Bologna, Italy \\
        \email{gabriele.columba@unibo.it}
        \and
        INAF – Osservatorio Astronomico di Padova, Vicolo dell’Osservatorio 5, 35122 Padova, Italy
        \and 
        School of Natural Sciences, Centre for Astronomy, University of Galway, Galway, H91 CF50, Ireland 
        \and
        INAF – Istituto di Radioastronomia, Via Gobetti 101, 40129 Bologna, Italy
        \and
        Univ. Grenoble Alpes, CNRS, IPAG, F-38000 Grenoble, France
        \and
        Université Côte d’Azur, Observatoire de la Côte d’Azur, CNRS, Laboratoire Lagrange, France
        \and
        Dipartimento di Fisica, Universit\`a degli Studi di Milano, Via Celoria, 16, Milano, I-20133, Italy
        \and
        European Southern Observatory, Alonso de C\'{o}rdova 3107, Casilla 19001, Vitacura, Santiago, Chile
        \and
        Institute of Astronomy, University of Cambridge, Madingley Road, Cambridge, CB3 0HA, UK
        \and
        Institute for Astronomy, University of Hawaii at Manoa, Honolulu, HI 96822, USA
        \and
        Instituto de Estudios Astrof\'isicos, Facultad de Ingenier\'ia y Ciencias, Universidad Diego Portales, Av. Ej\'ercito Libertador 441, Santiago, Chile
        \and
        Millennium Nucleus on Young Exoplanets and their Moons (YEMS)
    }

    \titlerunning{V721 CrA and BN CrA disks resolved in NIR}
    \authorrunning{Columba, G. et al.}
    
 
    \abstract
    {The environment within which stars form and evolve can play a crucial role in shaping their surrounding protoplanetary disks. This is the reason why homogeneous analyses of protoplanetary disks around young stars in the same star-forming region has become of great relevance in recent years. }
    {We present near-infrared scattered-light observations of the disks around two stars of the Corona Australis star-forming region, V721\,CrA and BN\,CrA, obtained with VLT/SPHERE in the $H$ band, as part of the DESTINYS large programme. Our objective is to analyse the morphology of these disks and highlight their main properties. 
    }
    {We adopted an analytical axisymmetric disk model to fit the observations and performed a regression on key disk parameters, namely the dust mass, the height profile, and the inclination. We used RADMC-3D code to produce synthetic observations of the analytical models, with full polarised scattering treatment.  
    } 
    {Both stars show resolved and extended disks with substructures in the near-IR. The disk of V721\,CrA is vertically thicker, radially smaller (\SI{\sim 120}{au}), and brighter than that of BN\,CrA (\SI{\sim 190}{au}). It also shows spiral arms in the inner regions. The disk of BN\,CrA shows a dark circular lane, which could be either an intrinsic dust gap or a self-cast shadow. 
    Both disks are compatible with the evolutionary stage of their parent subgroup within the CrA region: V721\,CrA belongs to the on-cloud part of CrA, which is dustier, denser, and younger, whereas BN\,CrA is found on the outskirts of the older off-cloud group. 
    }
    {}


\maketitle

\section{Introduction}
    \label{sec:intro}

    Studying populations of protoplanetary disks that formed in the same region is fundamental to understanding whether they share common properties across the sample (and possibly differences to other clusters) and what the cause of these similarities and contrasts is. 
    The environment around a disk can be distinguished between local and global. The former has to do with the immediate surroundings of the given star, including potential multiple companions \citep[e.g.][]{Moe&Distefano2017} and/or the proximity to ionising massive stars \citep[e.g.][]{Aru+2024}. The global environment can be intended as the star-forming region as a whole and its main macro-properties, such as the total mass in stars, the density, and the age, which can statistically impact the properties of its members \citep[e.g.][]{Manara+2023:ppvii} at the epoch of observation. 
    Eventually, the disk properties will affect planetary system architectures, but the link between the two statistics is still uncertain. 
    For this reason, large observing programmes such as the Disk Evolution Study Through Imaging of Nearby Young Stars  (DESTINYS, PI: C. Ginski) are targeting collections of disks from the same star-forming regions to reveal young protoplanetary disks with a homogeneous sample (see the surveys on the Chamaeleon, Taurus, and Orion regions by respectively \citealp{Ginski+2024:ChaI}, \citealp{Garufi+2024:Taurus}, and \citealp{Valegard+2024:Orion}).  
    
    In order to draw population trends, it is necessary to thoroughly characterise the single objects that are part of the given star-forming region. 
    The study of the two disks from Corona Australis (CrA) described in this paper responds to this exact purpose. The results and conclusions obtained from the following analysis will be important pieces of a larger effort \citep{Rigliaco2025} to characterise the properties of the CrA region, with a  framework similar to that of the study of \cite{Ginski+2024:ChaI}. 
    
    CrA is a molecular cloud known to be a formation site for low-mass stars \citep{Wilking+1985:CrA, Wilking+1992:CrA}.  
    The dust disks around CrA stars are on average less massive than coeval young clusters, as found by \citet{Cazzoletti+2019:CrA} through an Atacama Large Millimeter/submillimeter Array (ALMA) survey of this region, which cannot be attributed to the stellar mass function or photoevaporative effects.
    It is one of the closest star-forming regions to the Solar System, centred around \SI{150}{pc} \citep{Galli+2020:CrA, Zucker+2020}. 
    Recently, \citet{Galli+2020:CrA} and \citet{Galli+2022:poster} revised and expanded the CrA star census, combining Gaia DR2 \citep{GaiaCollab2018:GDR2} and EDR3 \citep{GaiaCollab2021:GeDR3} data with astrometric and photometric data delivered by the Cosmic-DANCe project \citep{Bouy+2013}. 
    In \citet{Galli+2020:CrA}, the authors reveal the existence of an older and more dispersed subgroup of cluster members, located in the northern part of the CrA dark cloud complex. They called these stars off-cloud, in contrast to the on-cloud members, which are instead closer to the main molecular clouds of CrA. 
    They found these two subgroups to have some differences in terms of proper motions, mean distance, and accretors' occurrence rate. Specifically, the off-cloud group appears to be centred around \SI{\sim 147.9 \pm 0.4}{pc}, against a distance of \SI{\sim 152.4 \pm 0.4}{pc} for the on-cloud stars. The latter also shows a higher frequency of accretors and the youngest stars of the CrA sample, indicating an overall less evolved population. 
    They find median ages for the on- and off-cloud subgroups of \SI{5}{Myr} and \SI{6}{Myr}, respectively.
    In \citet{Galli+2022:poster} the authors conclude that although the two groups in CrA are spatially distinct, their three-dimensional velocities are consistent with a common origin. 

    \citet{Ratzenboeck+2023a:Sco-Cen} divided CrA members, based on Gaia DR3 data \citep{GaiaCollab2023:DR3}, between the Main, North and \enquote{Scorpio-Sting} subclusters. The latter is the oldest and closest to the Sun ($\sim$ \SI{14.5}{Myr}, \SI{134}{pc}) and bridges the core of CrA to the Sco-Cen cluster. 
    The Main and North substructures, instead, are found to overlap significantly with the on- and off-cloud groups of \cite{Galli+2020:CrA}, respectively. 
    On an even larger scale, \cite{Ratzenboeck+2023b} found that the CrA star-forming region is the youngest endpoint of a chain of clusters \SI{100}{pc} in length  with a well-defined age gradient \citep{Posch+2023:CrA}, which interestingly contextualises CrA into a wider local star formation history. 
    The age estimates for CrA Main and North in \citet{Ratzenboeck+2023b}, however, are older than those of \cite{Galli+2020:CrA}: they place them respectively at ages of \SI{8.5}{Myr} and \SI{11.9}{Myr}, and at distances of \SI{155}{pc} and \SI{149}{pc}. These age differences are likely due to improvements in Gaia data between DR2 and DR3 and to a systematic difference in isochrone-fitting models: \cite{Ratzenboeck+2023b} refer to PARSEC isochrones \citep[e.g.][]{Bressan+2012:parsec, Marigo+2017:parsec}, whereas \cite{Galli+2020:CrA} relied mostly on the evolution models of \cite{Baraffe2015A&A}. 
    These differences are not too surprising as age determination for very young stellar groups is famously difficult (see e.g. the discussion in \citealp{Squicciarini&Bonavita2022:MADYS}).

    \begin{table*}
        \centering
        \caption{ Main stellar parameters of the two targets.}
        \label{tab:star_data}
        \begin{tabular}{l | c c}  
            \toprule
             Parameter &   V721 CrA    &   BN CrA         \\           
            \midrule
            Gaia DR3       & 6719035052475592576  & 6729805803970686592          \\
            Mass [$\Msun$]     &  1.0 $\pm$ 0.1\tablefootmark{a,b}   &  1.1 $\pm$ 0.1\tablefootmark{b}       \\
            Distance [pc]   & 156.85\tablefootmark{c} &   147.88\tablefootmark{c}         \\
            Age [Myr]       & 1.7\tablefootmark{b} -- 2.0\tablefootmark{d}    &   5.1\tablefootmark{e} -- 7.1\tablefootmark{b}   \\
            Spectral type      & K6IVe\tablefootmark{f}  &  K0\tablefootmark{g}-K9\tablefootmark{b} \\
            $ T_\mathrm{eff} $ [K]  &   4200\tablefootmark{b}    & 4400\tablefootmark{b} \\
            L [$\Lsun$]     &  0.80 $\pm$ 0.12\tablefootmark{b}      & 0.66 $\pm$ 0.02 \tablefootmark{b,e} \\
            R.A. [J2000]    & 287.441362\tablefootmark{c} & 279.1157549496\tablefootmark{c}     \\
            DEC. [J2000]    & -37.07405919\tablefootmark{c} & -39.0490765204\tablefootmark{c}     \\
            \bottomrule
        \end{tabular}
        \tablebib{ 
            (a) \cite{Woelfer+2023};
            (b) This work;
            (c) \citet{GaiaCollab2023:DR3};
            (d) \cite{Parker+2022};
            (e) \cite{Capistrant+2022};
            (f) \cite{Torres+2006};
            (g) \cite{Cieslinski+1998}. }
    \end{table*}

    The two disks that we analysed for this study, around V721\,CrA and BN\,CrA, are not well characterised in the literature. 
    This highlights the importance of the DESTINYS programme observations in expanding the sample of known disks with novel targets and to improving  the characterisation of close-by regions of star formation. 
    Studying in detail the single objects is crucial in order to draw reliable population trends. 
    We specifically chose these two disks among the DESTINYS targets of Corona Australis in virtue of their bright and extended structures in the near-infrared (NIR), which are best suited for a morphological characterisation. 
    
    V721\,CrA was the target of ALMA observations in 2015 (project ID: 2015.1.01301.S, PI: J. Hashimoto), and its band 6 data were analysed by \cite{Francis&vanDerMarel2020}. They fit for the first time a ring feature, resolved from the millimetre data, to a position angle (PA) of \ang{\sim 107} from the north and infer an inclination of \ang{\sim 55} to the line of sight (LoS), with a conservative uncertainty of \ang{5} for both angles. The millimetre-dust ring is at \ang{;;0.42} from the central star (or \SI{\sim 66}{au}) and is responsible for a flux of \SI{89}{mJy}. \cite{Woelfer+2023} expanded on the analysis of V721 with ALMA, analysing the gas kinematics in a collection of disks through bands 6 and 7. They present the moment 0 and 1 maps obtained from the emission of the \element[][12]{CO} and $^{13}$CO 2-1 line and argue that this system has a deep gas cavity and is subject to cloud absorption. 

    The disk of BN\,CrA has not been spatially resolved to date. It was chosen from among the targets of DESTINYS ESO programme ID 1104.C-0415(D) (PI: C. Ginski) owing to the infrared excess inferred from wide-sky photometric surveys, which hinted at the plausible presence of circumstellar dust. Its optical spectrum from \cite{Cieslinski+1998} shows a clear \Ha{} emission line and is compatible with a spectral type G7-K0.  

    The paper is organised as follows. In \secname~\ref{sec:data} we present the observational results on stars and disks. In \secname~\ref{sec:methods} we illustrate the parametric disk modelling, and in \secname~\ref{sec:results} we report on the numerical results. Finally, we discuss our results in \secname~\ref{sec:discuss} and conclude in \secname~\ref{sec:CrA_conclude}.

\section{Observational results}
    \label{sec:data}

    In this section we describe how we calculated consistently the stellar properties of both targets and describe the appearance of their disks from our new NIR images obtained with the Spectro-Polarimetric High-contrast Exoplanet REsearch (SPHERE) facility \citep{Beuzit+2019:SPHERE}.

    \subsection{Stellar properties}
        \label{subsec:star_props}

    The stellar properties were recalculated following the standard method highlighted by \citet{Garufi+2018:disks} and \citet{Garufi+2024:Taurus}. For both stars we obtained the effective temperature from the literature (see below) and used VizieR to construct the spectral energy distribution (SED). The stellar luminosity was then determined using a PHOENIX stellar photosphere model \citep{Hauschildt+1999}, scaled according to the Gaia distance and the R-band magnitude corrected for extinction, which was in turn derived from the $V$, $R$, and $I$ wavebands. The uncertainties in stellar luminosity were calculated by propagating errors from the distance, the extinction, and the spectral type. Finally, the stellar mass and age were estimated within the framework of isochrones, utilising a set of pre-main-sequence evolutionary models \citep{Siess+2000, Bressan+2012:parsec, Baraffe2015A&A, Choi+2016}.
    
    V721\,CrA (alternatively PDS 99 or 2MASS J19094592-3704261) is a K6IVe spectral type star \citep{Torres+2006} categorised with a high confidence level as a Class II object residing on-cloud \cite{Galli+2020:CrA}. We therefore adopted an effective temperature of 4200 K for the PHOENIX model. 
    The Gaia DR3 parallax \cite{GaiaCollab2023:DR3} translates to a distance of \SI{156.85}{pc}. 
    Our calculation of the visual extinction yields a value of 1.2 mag, while the stellar luminosity was \SI{0.80 \pm 0.12}{\Lsun}. These values yielded a stellar mass of \SI{1.0 \pm 0.1}{\Msun} and an age of 1.7--2.0\,Myr \citep{Rigliaco2025}.

    BN\,CrA (alternatively TIC 314944955 or 2MASS J18362779-3902561) was not included in the sample of \cite{Galli+2020:CrA}, but based on its galactic coordinates (longitude \ang{356}, latitude \ang{-14}), it is likely to be off-cloud, in the North subgroup. 
    The Gaia DR3 parallax of BN\,CrA corresponds to a distance of \SI{147.9 \pm 0.6}{pc}, which is closer to the average of the off-cloud members (i.e. CrA-North, at $149^{+7} _{-5}\,${pc}, \citealp{Ratzenboeck+2023a:Sco-Cen}). Its proper motion of $ \mu_\alpha^* = -1.22 \pm 0.03 \,\si{mas / yr} ,\,  \mu_\delta = -28.45 \pm 0.02 \,\si{mas \per yr} $ is also compatible with this group ($ \mu_\alpha^* = 1 \pm 2 \,\si{mas / yr} ,\,  \mu_\delta = -28 \pm 1 \,\si{mas \per yr} $, \citealp{Ratzenboeck+2023a:Sco-Cen}). 
    In this work we adopted an effective temperature of \SI{4400}{\K}. We determine a null extinction and a stellar luminosity of \SI{0.66 \pm 0.02}{\Lsun}. The  pre-main-sequence tracks therefore suggest a stellar mass of \SI{1.1 \pm 0.1}{\Msun} and an age of 5.1--7.1\,Myr \citep{Rigliaco2025}.

    \begin{figure}
        \centering
        \includegraphics[trim=0 45 0 52, clip, width=0.85 \columnwidth]{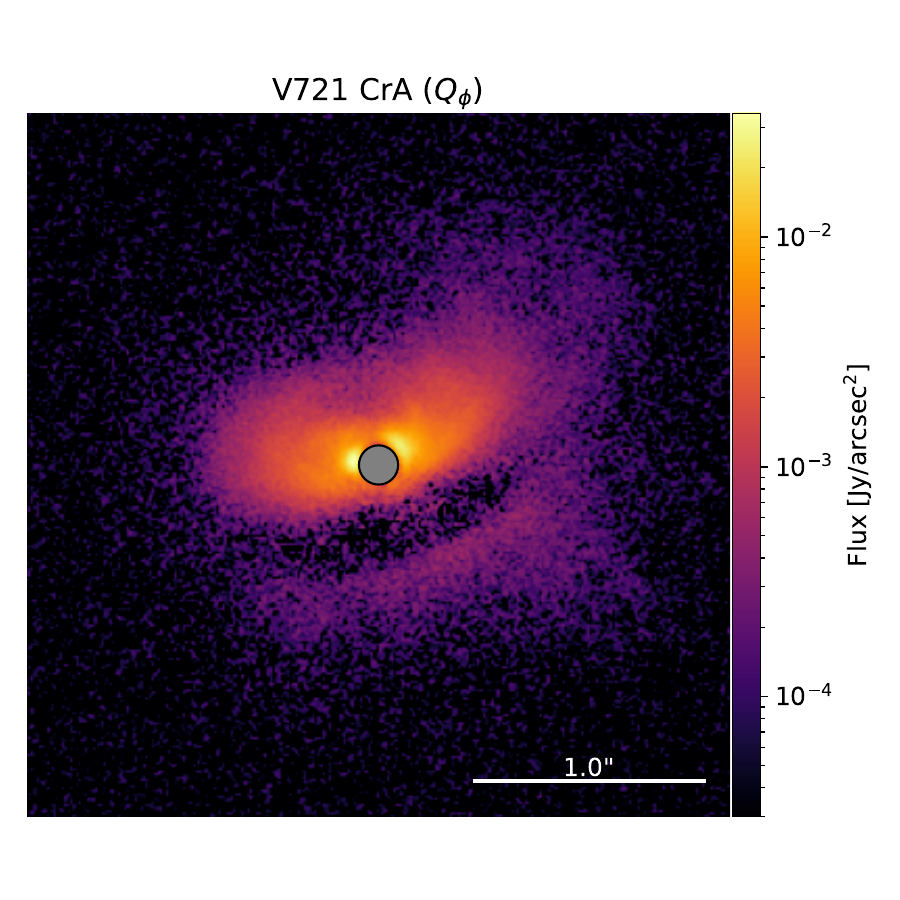}
         
        \includegraphics[trim=0 4.5cm 0 6.4cm, clip, width=0.85\columnwidth]{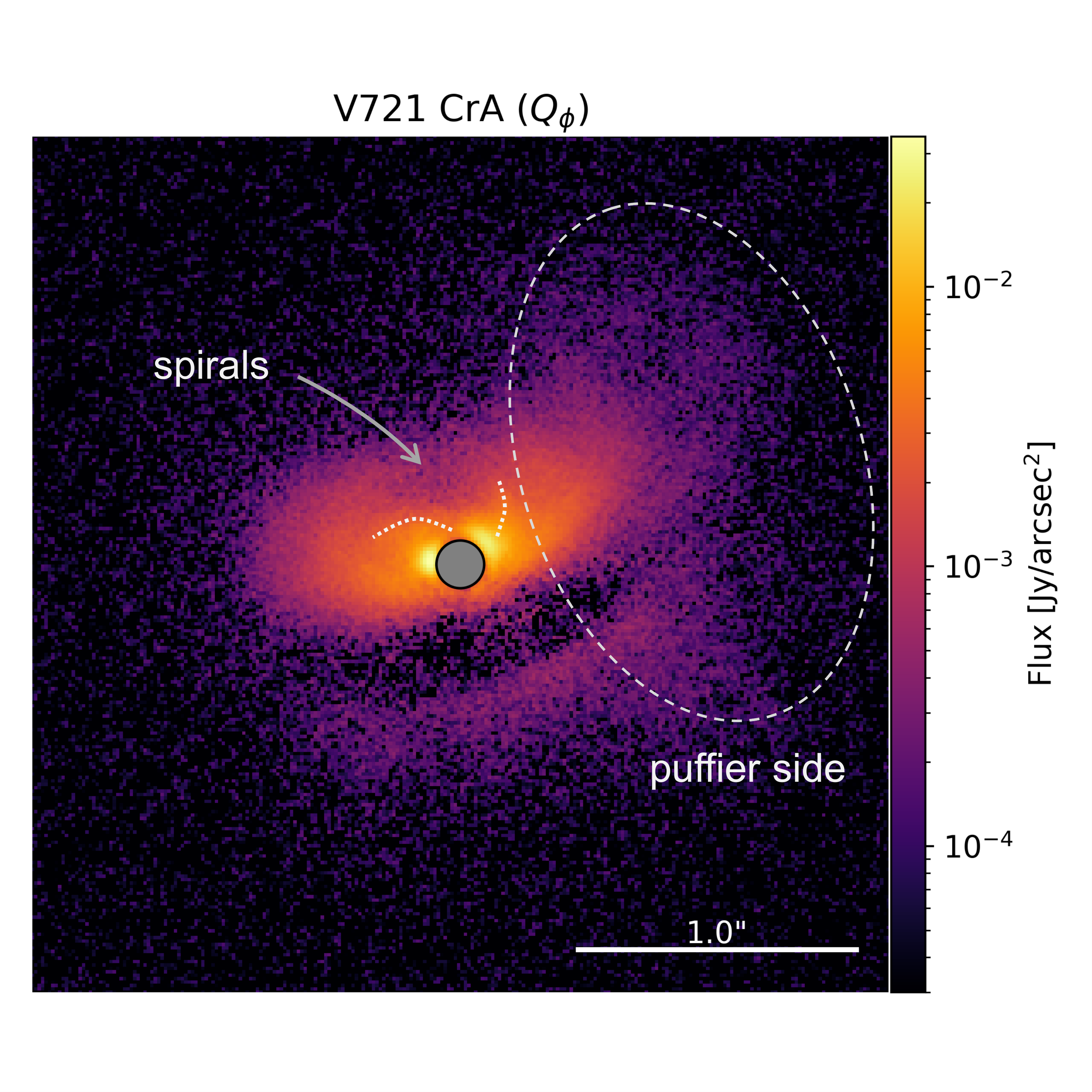}

        \caption{IRDIS $H$-band \Qphi{} frames for V721 CrA. 
        The colour scale is logarithmic and cuts out the read-out noise. The coronagraph area is covered by a grey filled circle; north is up and east to the left. 
        The bottom panel is the annotated version of the  top panel. The labelled spiral features  are more evident from the modelling residuals shown later in \figurename~\ref{fig:V721_mcmc-res} and \figurename~\ref{fig:V721_r2}. }
        \label{fig:V721_Qphi}
    \end{figure}

    \begin{figure}
        \centering
        \includegraphics[trim=0 45 0 52, clip, width=0.85\columnwidth]{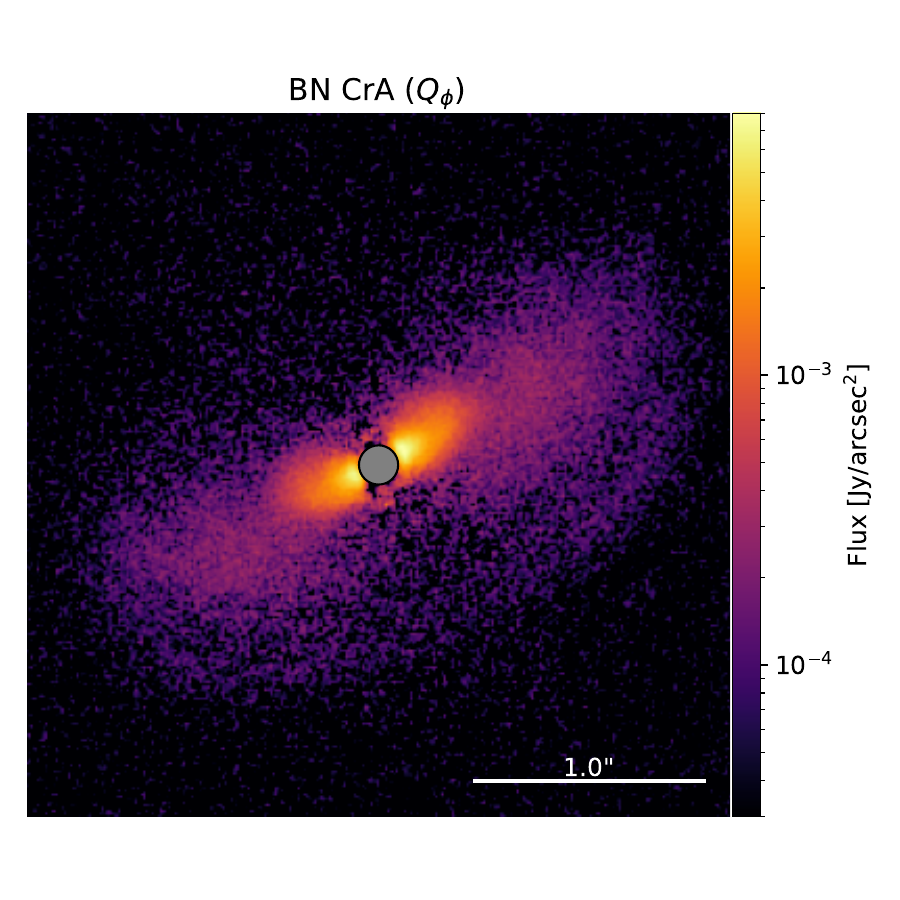}

        \includegraphics[trim=0 4.5cm 0 6.4cm, clip, width=0.85\columnwidth]{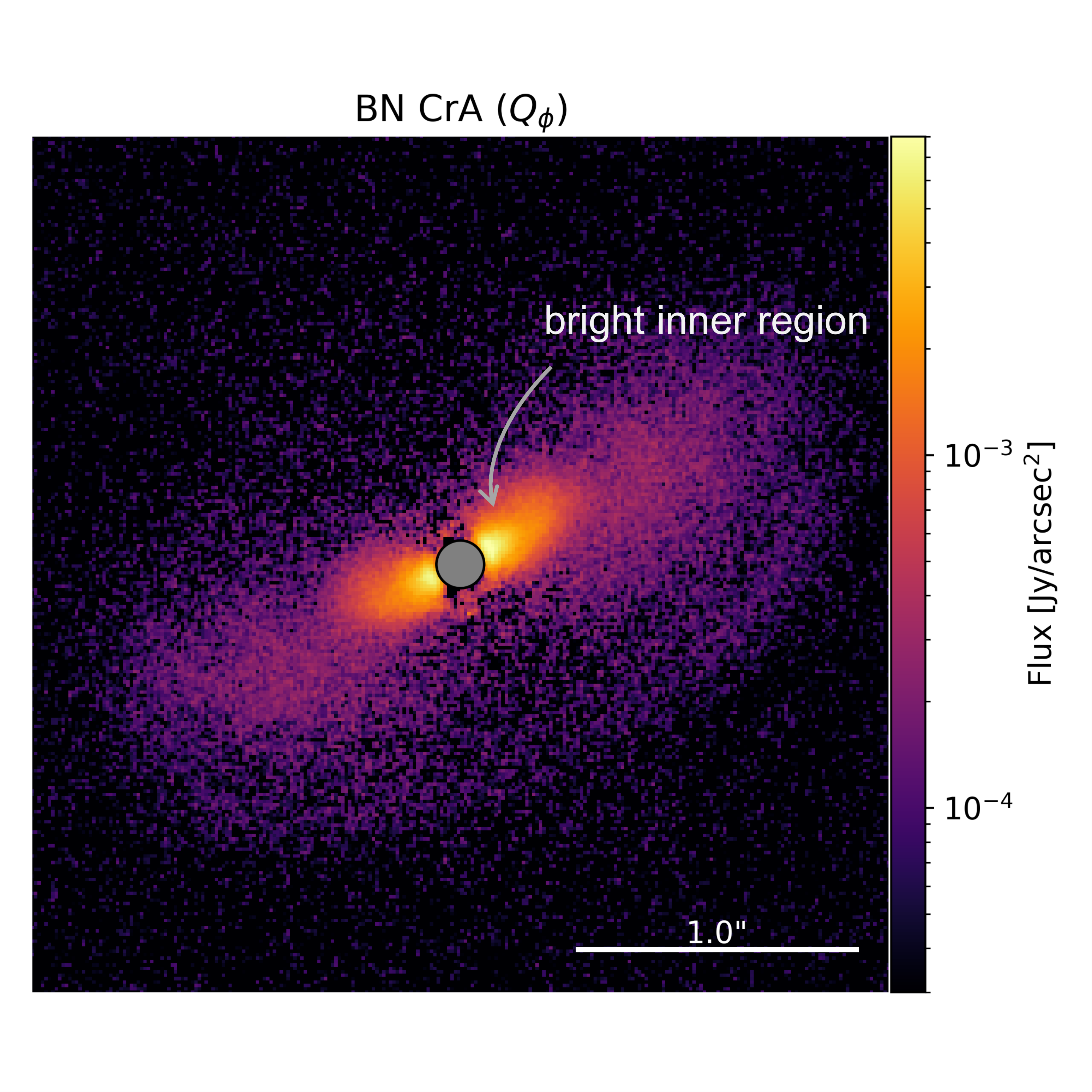}
        
        \caption{Same as \figurename~\ref{fig:V721_Qphi}, but for BN CrA. }
        \label{fig:BN_Qphi}
    \end{figure}

    \subsection{SPHERE observing set-up and data reduction}

    The ESO large programme DESTINYS, ID 1104.C-0415(D), targeted ten objects in the Corona Australis region to be observed in high-contrast imaging and polarimetry in order to resolve the dust-scattered light and inspect the disk structures. 
    In the following we describe the observations relative to the two disks analysed in this work, V721\,CrA and BN\,CrA (see \figurename~\ref{fig:V721_Qphi} and \figurename~\ref{fig:BN_Qphi}, respectively). 
    The observation epochs were 2022-07-01 and 2023-04-08, with effective total integration times of \SI{53}{min} and \SI{55}{min} (with \SI{32}{s} and \SI{64}{s} DITs), respectively, for V721\,CrA and BN\,CrA. 
    Thanks to the excellent seeing during both observation nights and the adaptive optics system of SPHERE, we report point spread functions with a full width at half maximum (FWHM) of \SI{\sim 55}{mas} for V721\,CrA and \SI{\sim 50}{mas} for BN\,CrA, very close to the theoretical diffraction limit. 
    The observation strategy and the data reduction methods were the same for both objects. 

    The disks were observed with the SPHERE instrument of the ESO Very Large Telescope (VLT) UT3. 
    In particular, the Infra-Red Dual-beam Imager and Spectrograph (IRDIS; \citealp{Dohlen+2008:irdis, Vigan+2010:irdis}) subsystem was used to target the NIR $H$-band at $ \lambda_0 \sim \SI{1.625}{\um}$ ($\Delta \lambda = \SI{0.290}{\um}$, pixel scale: \SI{12.251}{mas \per pix}  \citealp{Maire+2016:sphere}), in dual-beam polarimetric imaging mode \citep[DPI, see][]{deBoer+2020:irdisDPI, vanHolstein+2020:irdisDPI}. This passband has been successful in revealing the micron-sized dust scattering with high signal-to-noise ratios. 
    Polarimetry is also greatly suited for the study of circumstellar disks: central stars mostly emit unpolarised light, which can be easily distinguished from the polarised light scattered by dust instead. 
    
    Dedicated flux-calibration frames were acquired to allow conversion to physical flux units for the final data products and to ensure proper centring of the star. Specifically, the calibration frames consisted of short exposures of the target star with no coronagraph; from these, the total-intensity flux (or Stokes I frame) of the star was compared to its tabulated magnitude in the 2MASS catalogue as photometric calibration of the photon counts. Lastly, additional corrections were applied to the polarised fluxes measured at the detector, as explained in detail in \citet{vanHolstein+2020:irdisDPI}, to also account  for instrumental polarisation effects along the optical path of SPHERE/IRDIS. Using this calibration technique, the final data products are not affected by Strehl-ratio issues that instead can arise from calibrating with photometric standard stars.
    
    The observations consisted of 14 polarimetric cycles, each containing four exposures taken at half-wave plate (HWP) switch angles \ang{0}, \ang{45}, \ang{22.5}, and \ang{67.5}. 
    The observing strategy was pupil tracking \citep{vanHolstein2017:irdis}, which allows rotation of the field of view (FoV) in time with respect to the detector. The N\_ALC\_YJH\_S coronagraphic mask with an inner working angle of \SI{92.5}{mas} was used for the scientific frames to suppress the central starlight. 
    The data were reduced with the IRDAP v1.3.4 tool\footnote{\url{https://irdap.readthedocs.io/en/latest/}} \citep{vanHolstein+2020:irdisDPI}.
    The default settings were adopted for this reduction to apply polarimetric differential imaging (PDI; \citealt{Kuhn+2001}) and produce the frames corresponding to the Stokes parameters $I$, $Q$, $U$, \Qphi{}, and $ U_\phi $. 
    The azimuthal Stokes parameters \Qphi{} and $U_\phi$ were calculated according to the definitions of \citet{deBoer+2020:irdisDPI}:
        \begin{equation}
            \begin{array}{@{}l@{}}
                Q_\phi = - Q\cos( 2\phi ) - U\sin( 2\phi ), \\
                U_\phi = Q\sin( 2\phi ) - U\cos( 2\phi ). \\
            \end{array}
        \end{equation} 
    Here $\phi$ is the east-of-north (EoN) polar angle in the coordinate system centred on the target star. 
    The \Qphi{} is the frame where the dust signal is best recovered against the stellar light, and this is of central importance in our subsequent analysis. 
    The $U_\phi$ frame, on the other hand, is related to a non-axisymmetric polarisation, which is often interpreted as a signal resulting from multiple scattering events, perhaps out of thicker disk regions. 
    We report for completeness all four polarimetric Stokes frames in the Appendix, for V721\,CrA in \figurename~\ref{fig:V721_quadStokes} and for BN\,CrA in \figurename~\ref{fig:BN_quadStokes}. 
    
    We reduced the Stokes frames of each disk with a dedicated approach aimed at optimising the subtraction of the stellar contamination from the dust polarised signal. 
    This proved to have a tangible impact on the final science products, as a spurious signal was present for both objects; specifically, in the form of a blob inside the dark lane for V721\,CrA and as an axial over-brightness for BN\,CrA (aligned along the minor disk axis, deceiving itself as a potential jet). 
    The optimal stellar polarisation to subtract was found with a methodology already adopted in previous works as \citet{Haubois+2023} and \citet{Garufi+2024:Taurus}. Namely, with a grid-search approach for V721\,CrA and minimising the negative signal in the \Qphi{} for BN\,CrA. We thus report a degree of linear polarisation from the two stars of 0.79\,\% and 0.63\,\%, respectively.

    \subsection{SPHERE images description}

        The disk of V721\,CrA appears as a quite thick and flared disk from scattered light (\figurename~\ref{fig:V721_Qphi}): the coronagraph spot almost grazes the closest disk edge, whereas the other half of the disk surface is more exposed. A thick midplane dark lane also reveals the three-dimensional geometry of this disk and its orientation. 
        The disk is overall axisymmetric and regular, with a few exceptions. Firstly, the western side appears puffier and more extended than the other half, as one can see from the greater height of the scattered signal and a slimmer midplane dark lane. Then, there are some substructures within the disk resembling spiral wakes, developing in the radial direction. 
        The inner regions at the border of the coronagraph are bright, and there is no evident resolved cavity and no projected shadow on the disk surface.  
        Some disk substructures are highlighted with a higher contrast in the model residuals, described in \secname~\ref{subsec:V721_results} for V721\,CrA. 
    
        BN\,CrA displays a wide and axisymmetric disk in scattered light (\figurename~\ref{fig:BN_Qphi}). The disk consists of a bright inner region and a more tenuous outer region. 
        Likely due to the polarised scattering properties, the farther side of the disk is so faint in the outer region that it is barely detected above the noise. 
        The disk of BN\,CrA appears to have a thinner aspect ratio and a less evident midplane dark lane compared to V721\,CrA. 
        The disk substructures are highlighted with higher contrast in the model residuals image, described in \secname~\ref{subsec:BN_results} for BN\,CrA.

\section{Numerical methods}
    \label{sec:methods}

    In the following subsections we illustrate the methods used to analyse the two disks, based on the observed data described in \secname~\ref{sec:data}. 
    The main goal of this analysis was to characterise the new disk structures resolved for the first time in NIR scattered light. 
    Given the general similarities between the disks of V721\,CrA and BN\,CrA, we were able to apply similar techniques for their characterisation.  
    Firstly, the PAs of the disks in the sky plane were computed by fitting ellipses to the intensity contours, as described in \secname~\ref{subsec:CrA_contourfitting}. 
    Then, to estimate several basic disk properties at once, we performed a Markov chain Monte Carlo (MCMC) regression of the data using a simple parametric disk model (see \secname~\ref{subsec:CrA_parammodelling}). We used RADMC-3D v2.0\footnote{
    \url{https://github.com/dullemond/radmc3d-2.0}
    } \citep{Dullemond+2012:RADMC3D} to simulate the scattered light synthetic observation of our disk models, to then compare with the data. 
    Finally, we extracted the scattering phase functions from the data and the relative best-fit models.

        \begin{figure*}
            \centering
            \begin{subfigure}[c]{0.42\textwidth}
                \centering
                \includegraphics[trim=0 45 0 20, clip, width=\textwidth]{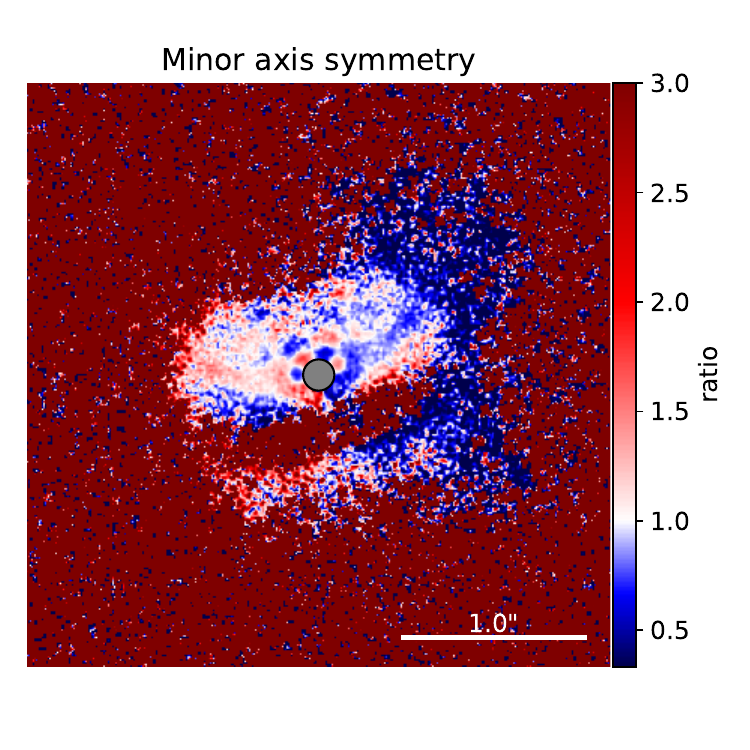}
                \caption{V721 CrA}
                \label{fig:V721_specularity}
            \end{subfigure}
            \qquad 
            \begin{subfigure}[c]{0.42\textwidth}
            \centering
                \includegraphics[trim=0 45 0 20, clip, width=\textwidth]{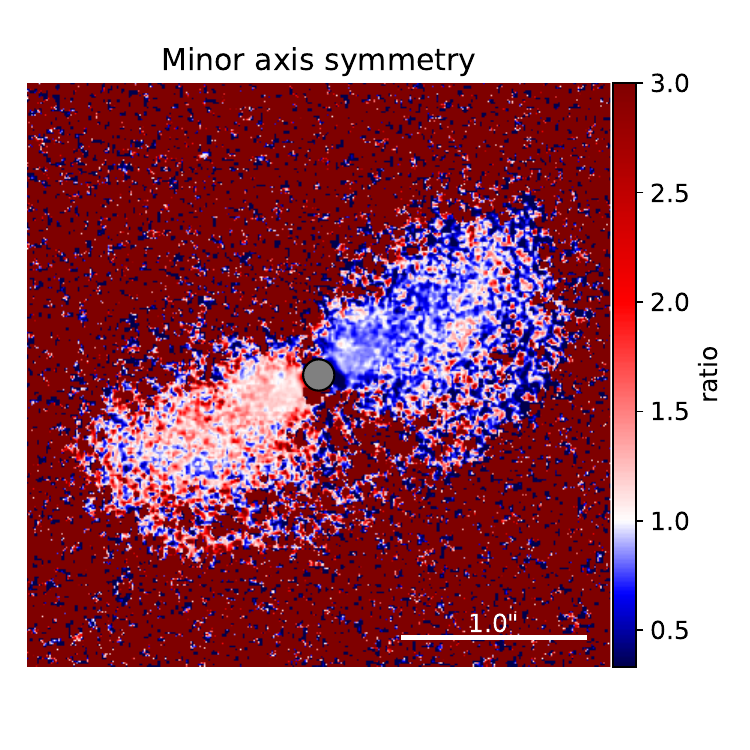}
                \caption{BN CrA}
                \label{fig:BN_specularity}
            \end{subfigure}
            
            \caption{ Brightness ratio around the minor disk axis in the \Qphi{} frames for V721\,CrA and BN\,CrA (the latter after a Gaussian smoothing with a 0.5 pixel sigma). The mirror axis inclination is orthogonal to the best PA values reported in \secname~\ref{subsec:CrA_contourfitting} (namely, \ang{104.4} for V721\,CrA and \ang{117.5} for BN\,CrA, east-of-north to the disk major axis). 
            The coronagraph area is covered by a grey circle. 
            North is up and east to the left. }
            \label{fig:CrA_specularity}
        \end{figure*}

    \subsection{Preliminary contour fitting}
        \label{subsec:CrA_contourfitting}

        The two disks around V721\,CrA and BN\,CrA are extended and mostly axisymmetric. 
        We measured the extension of the disks directly on the \Qphi{} frames, illustrated in \figurename~\ref{fig:V721_Qphi} and \figurename~\ref{fig:BN_Qphi}, along the disk major axes, using the pixel scale of IRDIS of \SI{12.251}{mas/pix} and Gaia DR3 parallaxes to each object to convert to au. 
        We roughly estimate an outer radius $r_\mathrm{out}$ of \SI{\sim 120}{au} for V721\,CrA and \SI{\sim 190}{au} for BN\,CrA. The boundaries of disks in scattered light are not particularly sharp, and thus these estimates have an uncertainty that can be around 5--10\% of the $r_\mathrm{out}$ itself. 
        The brighter inner part of the BN\,CrA disk extends to around \SI{80}{au}, one-half of the total extension. We do not find a similar knee in the radial brightness of V721\,CrA. 

        To determine the mean PA of the disks from the scattered light, we relied on brightness contour fitting. 
        Considering the general symmetry of the disks, the signal intensity should be approximately constant at a given radial distance from the central star. Since both disks are seen at significant inclinations, a constant-radius circle becomes an ellipse with a clear semi-major axis orientation. 
        Then, we fitted an ellipse to each brightness contour level to find the PA of the disk, using ten levels to average over the whole disk surface. 
        We selected the levels on a 3 pixel smoothed \Qphi{} frame, ensuring that the minimum level was not excessively distorted by the background noise. We also imposed having more than 20 points to fit a level, and we discarded all the points behind the coronagraph area or belonging to the dust below the disk midplane. 
        With these conditions, we fit each contour level with the least-squares ellipse fitting algorithm by \cite{Hammel&Molina2020}.\footnote{ \url{https://github.com/bdhammel/least-squares-ellipse-fitting/tree/v2.0.0} } 
        Finally, we computed the best-fit value as the median of each ellipse parameter among the ten levels, with the standard deviation as the uncertainty. 
        The best PA results are \ang{104.4 \pm 1.9} for V721\,CrA and \ang{117.5 \pm 1.0} for BN\,CrA. 
        The inclination can also be roughly estimated from the ratio of the ellipse axes; the best-fit results are \ang{\sim 63} and \ang{\sim 72}, respectively, for the two disks. However, an important caveat is that, due to the polarised scattering phase function, the observed brightness is lower along the projected disk minor axis. Thus, an ellipse fitting of brightness contours will be biased to suggest higher inclinations than the real ones. 

        Another check on the PA was performed on the basis of the symmetry of the disk with respect to the minor axis. Basically, we computed the average mirror disk image as the mean between the original data and the data flipped around a specular axis with a given PA. We then calculated the residuals between the mirror average and the original data. 
        For an axisymmetric disk, the residuals should be minimised for a mirror axis with PA corresponding to the disk PA. 
        We find that with this procedure we obtain PA values compatible with the previous ones resulting from the ellipse fit within a margin of \ang{1}. 
        To visualise the brightness asymmetries of the two disks, we illustrate in \figurename~\ref{fig:CrA_specularity} the ratio of the \Qphi{} frames to their copy flipped around the minor disk axis. 
        From these images, it is possible to see that V721\,CrA has a more complex pattern of brightness asymmetries;   the most striking feature is an extended structure on the north-western edge. BN\,CrA, on the other hand, presents a distinct left--right asymmetry, but with lower ratios overall.

    \subsection{Parametric disk model}
        \label{subsec:CrA_parammodelling}

        Understanding the morphology of a disk is not a trivial task, and it involves numerous different variables. Moreover, analysing it based on the scattered light signal alone carries its own limitations, due to many uncertainties and possible degeneracies, which can lead to misinterpretation of a disk feature. 
        Nevertheless, it is important to characterise every object as much as possible with the data at hand, being aware of the limitations and the approximations (which are necessary). 

        In order to infer a few basic properties of our disks, we decided to adopt an axisymmetric analytical parametric model and to perform regressions to obtain the best-fit parameters. 
        Our model does not claim to be complete, as that goes beyond the scope of this work; instead, the model focuses mostly on the observable that we are able to constrain, based on our observed data: the NIR scattered light. 
        Thus, our model looks at the small dust component of the disk (\SI{\sim 0.1}{\um}) responsible for the polarised light scattering, mainly off the surface layers. 
        For this reason, we can neglect the larger millimetric dust grains on the midplane and the temperature profile of the disk. 
        The parameters targeted by this analytic model are low dust mass \Md{}, inclination $i$, and dust density profile in terms of reference scale height $ h_0 $ and flaring exponent $\beta$. 
        The dust density of our model follows the standard gas density for a vertically isothermal disk in hydrostatic balance (see e.g. \citealp{Andrews2015:ppds, Bae+2023:PPVII, vanderMarel&Pinilla2024}), with the caveat that we leave the scale height profile to be parametrised by a simple geometric form. This is considered a good approximation for the micron-sized dust, which is known to be strongly coupled to the gas, and indeed it has been used in similar previous works, for example \citet{Tazaki+2023:IMLup}. 
        The dust density, as a function of the cylindrical coordinates $ (r, z) $ (the radius and the height over the midplane), then assumes the  form 
        \begin{equation}    \label{eq:rho_dust}
            \rho_\mathrm{d} (r, z) =  \frac{ \Sigma_\mathrm{d} (r) }{ \sqrt{2 \pi h_\mathrm{d}^2} } \cdot \exp{ \left[ -\frac{ z^2 }{2 h_\mathrm{d}^2} \right] } \ ,
        \end{equation}
        where $ \Sigma_\mathrm{d} $ is the radial dust surface density and $ h_\mathrm{d} $ is the dimensional dust scale height. They are assumed to obey, respectively,
        \begin{equation}    \label{eq:Sigma_dust}
            \Sigma_\mathrm{d} (r) = \Sigma_\mathrm{0} \left( \frac{r}{r_0} \right)^{-1} ,
        \end{equation}
        \begin{equation}    \label{eq:h(r)}
            h_\mathrm{d} (r) = \frac{h_0}{r_\mathrm{0}} r \left( \frac{r}{r_0} \right) ^{\beta}  \ .
        \end{equation}
        The surface density $ \Sigma_\mathrm{d} $ is normalised to sum up to \Md{} on the whole disk. 
        In the remainder of the paper we simplify the notation of the reference aspect ratio $h_0 / r_0$ as $hr_0$, which indicates the variable's value at the reference radius $r_0$, which we chose to be \SI{115}{au}. This is arbitrary and has no impact on the final shape of the model. 
        Depending on the scale height formulation, the flaring index can be found in the value $ ( 1 + \beta) $, and in any case $ \beta > 0 $ to have an illuminated disk.
        These parametric models extend radially from $ r_\mathrm{in} = \SI{1}{au} $ to $ r_\mathrm{out} = \SI{120}{au} $ for V721\,CrA and to $ r_\mathrm{out} = \SI{190}{au} $ for BN\,CrA, as determined in the preliminary analysis described in the previous subsection. The inner radius $ r_\mathrm{in} $ is also arbitrary, as the coronagraph covers the innermost regions of both disks, thus making it impossible for us to constrain \rin{} through image analysis.

        \subsubsection*{RADMC-3D set-up}
        We used RADMC-3D (v2.0) to generate our model numerically and the relative scattered-light synthetic observations. 
        The model grids have 200 logarithmically spaced radial cells from \rin{} to \rout{} and 200 uniformly spaced cells for the zenith angle $\theta$ in the interval $[\ang{30}, \ang{150}]$. 
        Since our model is axisymmetric, the azimuthal dimension $\varphi$ could be deactivated, leaving in practice a 2D $ (r-\theta)$ grid repeated for all $2\pi$ of the disk plane. RADMC-3D, in this regard, has a special treatment of axisymmetric grids that is computationally efficient with anisotropic scattering. We noted that setting a lower number of scattering photons with axisymmetry gave better results and a faster run time than with a 3D grid and a thousand times more photons. We limited the precision of azimuthal resolution using the flag \texttt{dust\_2daniso\_nphi=90}. 
        For the synthetic images, we imposed full anisotropic polarised scattering treatment, with $10^5 $ photon packages, to be able to simulate the \Qphi{} signal from our model. As anticipated, we neglected the thermal photon scattering, as we were not interested in the temperature profile of the disk. 
        Moreover, the scattered-light NIR signal is not significantly affected by the temperature of the dust grains in most of the disk where the temperatures are low (everywhere outside the coronagraph). 
        To further speed up the RADMC-3D computations, we reduce the default sub-pixelling settings by using the \enquote{\texttt{sloppy}} flag when creating the synthetic image. 
       
        The PA of the disk model in the image was fixed to be the best-fit PA from the preliminary analysis described in \secname~\ref{subsec:CrA_contourfitting}. 
        Finally, we applied Gaussian smoothing to the simulated Stokes frames to convolve the model to the observational angular resolution. Specifically, we calculated a convolution scale $ \mathrm{FWHM} = 1.025 \cdot \lambda_0 / D = \SI{42.4}{mas}$, based on the IRDIS pixel scale and the $H$-band diffraction limit at the VLT ($D = \SI{8.1}{m}$ primary mirror).

        \subsubsection*{Dust opacities}
        
        The dust scattering properties, or \enquote{opacities}, are the key ingredient to go from density distribution to final synthetic images. This is particularly important when dealing with polarised light: the scattering coefficients are dependent on the photon's incoming angle, and this is true for each scattering event. 
        This requires specifying the dust grain coefficients for all possible angles and all grain sizes to be able to compute images with full-Stokes scattering treatment. 
        The basic disk modelling with these parameters was already too computationally intensive to leave dust opacity as another free parameter. Thus, we performed only a few test runs to decide which dust species to use, comparing the DIANA standard dust opacities \citep{Woitke+2016:diana} to the default RADMC-3D silicate grains. We produced the DIANA opacities through OpTool \citep{Dominik+2021:optool} as an average of a collisional distribution of grains between \SIrange{0.1}{50}{\um} composed by 87\% pyroxene and 13\% carbon, with a maximum volume porosity of 25\%. 
        The RADMC-3D default grains, labelled `silicates', instead consist of mono-dispersed grains of \SI{0.1}{um} made of amorphous olivine (50\% Fe and 50\% Mg). The scattering matrices for this material were computed with Mie scattering by C. Dullemond based on the optical constants of \cite{Jaeger+1994:silicates} and \cite{Dorschner+1995:silicate}. 
        Eventually, we adopted the silicate grains for our MCMC models, as we found that using DIANA opacities resulted in two dark stripes on the closer disk side, resulting in a poorer fit to the disk inclination and worse MCMC chains overall. We will leave a deeper study of the impact of dust properties for possible future works.

        \subsubsection*{MCMC modelling set-up}
        The disk model was included in a MCMC procedure with the goal of finding the best-fitting disk parameters. 
        We used the Python module \texttt{emcee} \citep{emcee2013} to perform our MCMC regression, comparing the disk observation with our synthetic model at each step, pixel by pixel. 
        The observation data were masked to improve the performance and to make sure that only the relevant signal was fitted. To this end, we masked with \enquote{NaNs} the areas behind the coronagraph or below the standard deviation of background noise, after applying a Gaussian smoothing to the \Qphi{} data. 
        The logarithmic likelihood was chosen to be a standard Gaussian distribution of the residuals between the model and the data, and we employed flat priors on our free parameters (details of the priors ranges are reported in \tablename~\ref{tab:priors}). 

        The running time for this MCMC regression was high due to the intrinsic full-scattering computation and the incompatibility of RADMC-3D model generation out-of-the-box with parallel emcee walkers on independent disk models. Due to time constraints  we had to limit our MCMC runs to $\sim$2000 steps and 32 walkers for each run. 
        We made sure to have at least a general consistency in the evolution of the ensemble of independent walkers along the whole chain. For our parameters, the chains do not show a significant evolution of the walker  distribution after the removal of the first hundred steps (see e.g. Appendix \figurename~\ref{fig:V721_chainsteps} the \enquote{post-burn-in} chains). Thus, we discarded the first 500 steps from the chains and computed the final best-fit values as the median of the marginalised posteriors. We deemed the final values extracted from the MCMC procedure physically realistic.

\section{Numerical results}
    \label{sec:results}

    We report in this section the main results for the two CrA disks, analysed as described in \secname~\ref{sec:methods}. 
    Complementary figures have been added to the Appendix to keep an easier readability of the main text.

    \subsection{V721 CrA results}
        \label{subsec:V721_results}

        \begin{figure}
            \centering
            \begin{subfigure}[c]{\columnwidth}
                \centering
                \includegraphics[trim=0 42 0 30, clip, width=\textwidth]{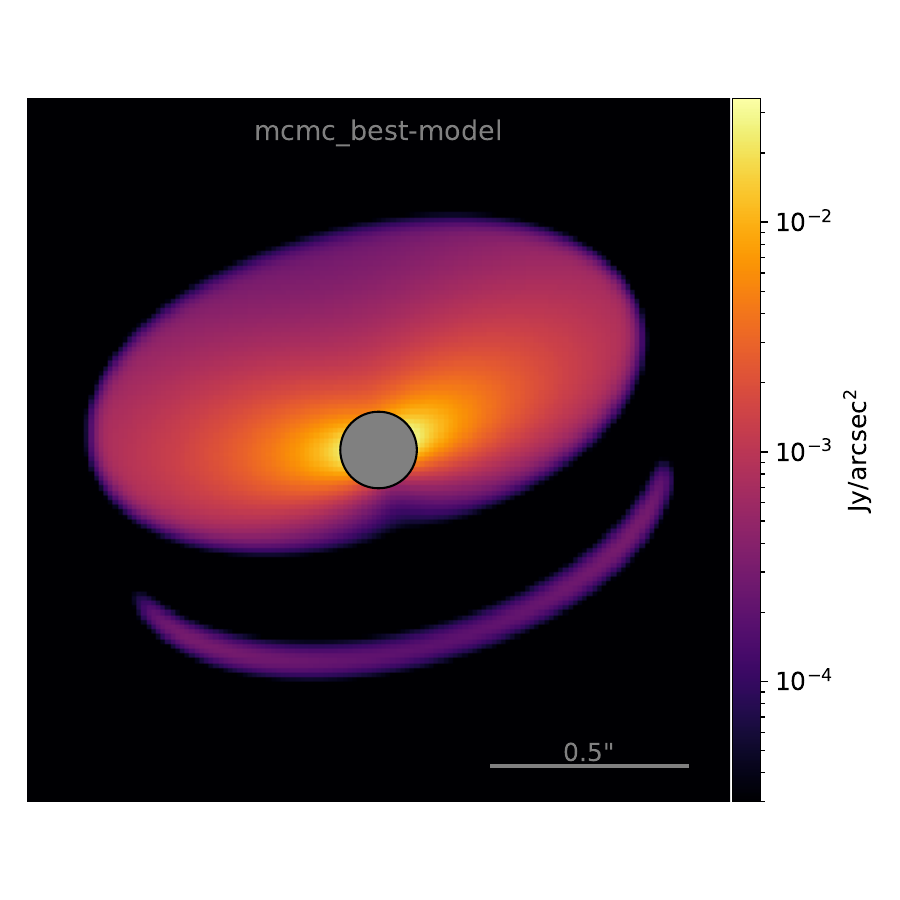}
                \caption{V721 CrA best model}
                \label{fig:V721_best-mcmc}
            \end{subfigure}
            
            \begin{subfigure}[c]{\columnwidth}
            \centering
                \includegraphics[trim=0 45 0 30, clip, width=\textwidth]{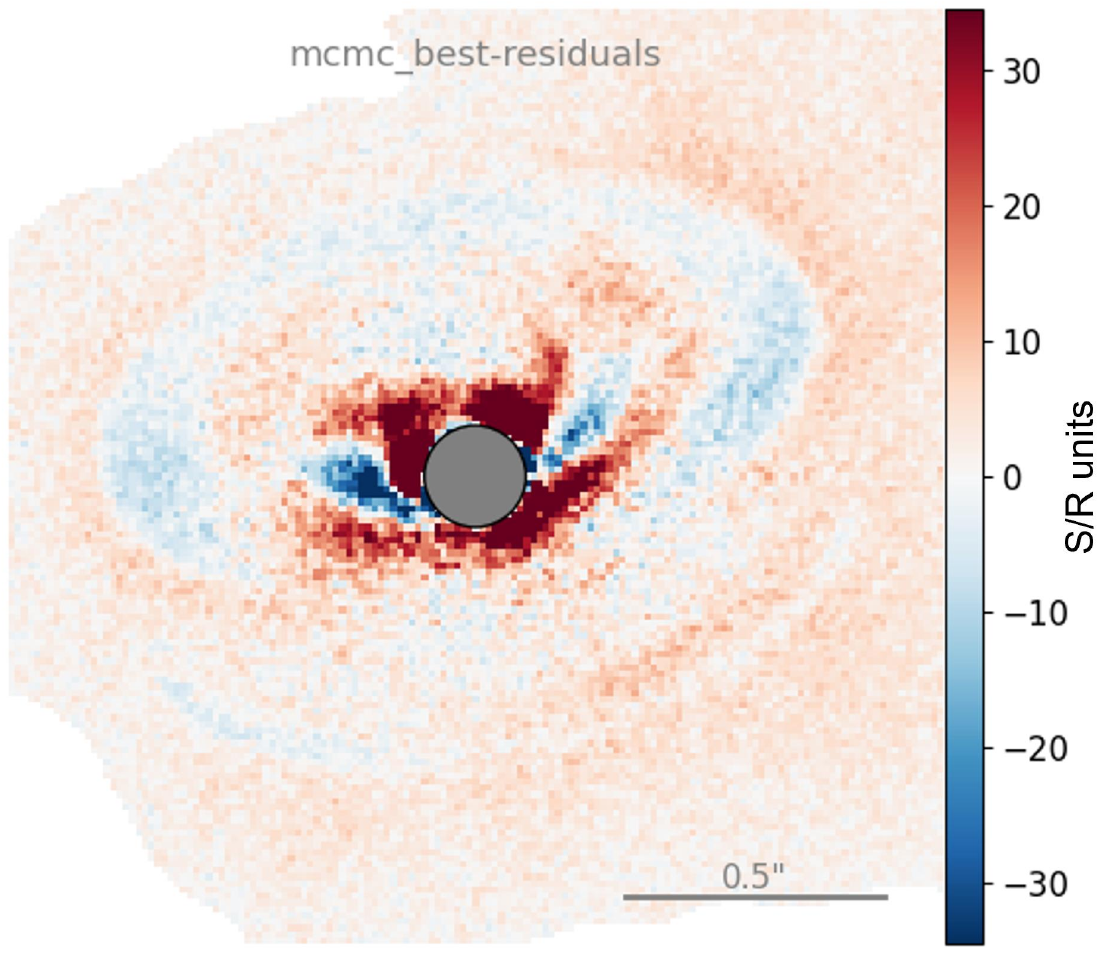}
                \caption{ Best model residuals}
                \label{fig:V721_mcmc-res}
            \end{subfigure}
    
            \caption{Best model for V721\,CrA resulting from the MCMC procedure. Panel (a) illustrates the best model at $\lambda = \SI{1.625}{\um}$, convolved to the data resolution. The grey circle represents the coronagraph. Panel (b) shows the residuals between the data and the best model. North is up and east is left.  }
            \label{fig:V721_mcmc}
        \end{figure}

        The disk around V721\,CrA is best matched by the model shown in \figurename~\ref{fig:V721_mcmc}, whose best-fit parameter values are reported in \tablename~\ref{tab:best-mcmc} along with their confidence intervals resulting from the MCMC sampling. 
        The aspect ratio at the reference radius of $ r_0 = \SI{115}{au}$ corresponds to  $ h/r \approx 0.1 $, in line with common values in samples of disks (see e.g. Fig.5 from \citealp{Avenhaus+2018:dartts}) and similar to the disks around IM Lup and MY Lup \citep{Avenhaus+2018:dartts}. 
        The flaring index, the exponent $ \beta $ in our \eqname~\ref{eq:h(r)}, is at $\sim 0.38 $. 
        The best-fit dust mass corresponds to \SI{2.8e-4}{\Msun}; we recall that we only modelled the micrometric dust. 
        The model for V721\,CrA is consistent overall with the young age attributed to this target and suggests that this disk still has abundant material from the formation. 
        The inclination of our best parametric model matches within \ang{\pm 3}) the inclination found by \cite{Francis&vanDerMarel2020} for the millimetre dust ring. This compatibility of the inclination between the millimetre ring and the scattered light structure suggests that the whole disk may have a well-defined midplane. 
        
        In the residuals of the best model of V721\,CrA (\figurename~\ref{fig:V721_mcmc-res}) there are a few features to note. Firstly, in the inner regions of the disk, two spiral arms appear with a strong S/N contrast above our model, winding clockwise; these are the same structures that are best visible in the flux-corrected maps, illustrated  in \figurename~\ref{fig:V721_r2}. 
        Then, it is possible to see that the outer edges of our model are brighter than the data: especially on the upper disk side, the residuals show a moderate over-subtraction. This suggests that the outer parts of the disk may locally deviate from the simple aspect ratio formulation of the model, perhaps due to an exponential decay in the dust density (which we did not include). 
        Lastly, the residuals show a remarkable lack of signal near the coronagraph, along the major disk axis. This might correspond to the carving of a cavity in the space close to the spiral arms, depleting the inner disk first and thus lowering the observed brightness.

        \begin{table}[bt]        
            \centering
            \caption{Best-fit disk parameters obtained from the MCMC posteriors.  }
            \renewcommand{\arraystretch}{1.4} 
            \setlength{\tabcolsep}{11pt}
            \begin{tabular}{l | c c }
                \toprule
                 Parameter &   V721 CrA   &   BN CrA    \\
                \midrule
                $ M_\mathrm{d} $ [\si{\Msun}]  &    2.8e-04$^{+1.8e-04} _{-1.3e-04} $   &  1.0e-06$^{+1.6e-07} _{-1.1e-07} $    \\ 
                $ hr_0 $   &   0.095 $^{+0.006} _{-0.004} $   &   0.070 $^{+0.009} _{-0.005} $      \\ 
                $ \beta $   &   0.38 $^{+0.01} _{-0.01} $    &   0.42 $^{+0.02} _{-0.04} $     \\ 
                $ i $ [°]  &  57.6 $^{+0.3} _{-0.2} $  &  70.6 $^{+0.7} _{-0.6} $ \\ 
                \bottomrule
            \end{tabular}
            
            \label{tab:best-mcmc}
        \end{table}

    \subsection{BN CrA results}
        \label{subsec:BN_results}
        
        \begin{figure}[t]    
            \centering
            \begin{subfigure}[c]{\columnwidth}
                \centering
                \includegraphics[trim=0 42 0 20, clip, width=\textwidth]{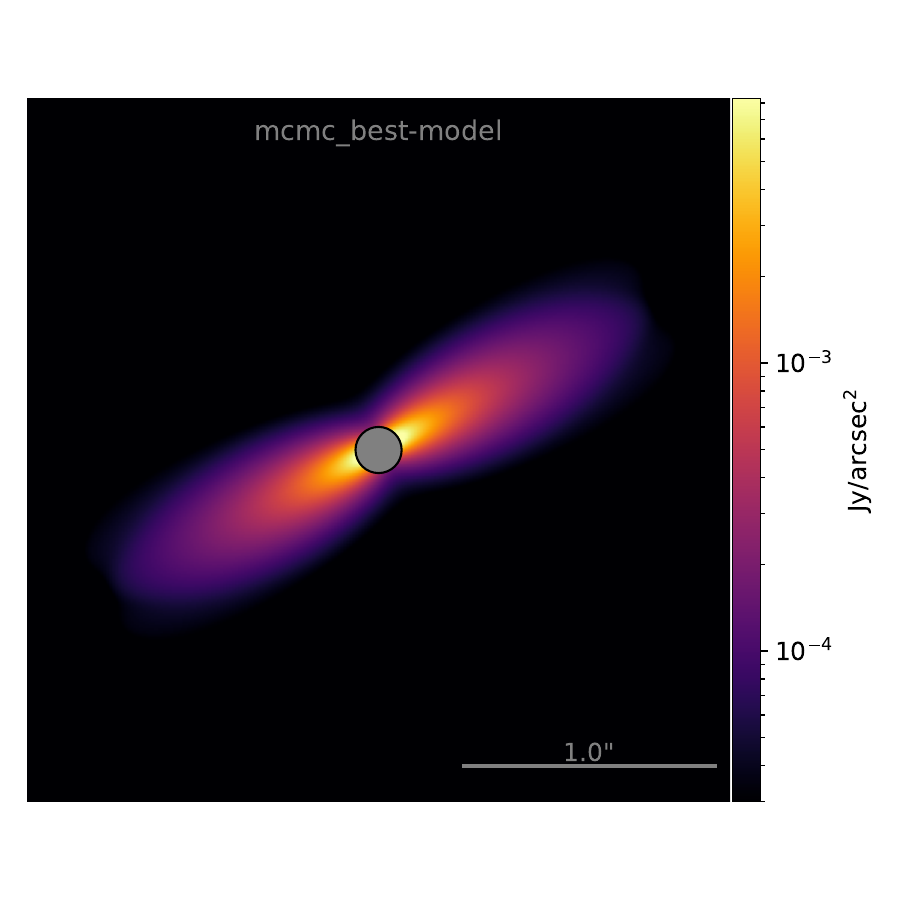}
                \caption{BN CrA best MCMC model}
                \label{fig:BN_best-mcmc}
            \end{subfigure}
            
            \begin{subfigure}[c]{\columnwidth}
            \centering
                \includegraphics[trim=0 45 0 20, clip, width=\textwidth]{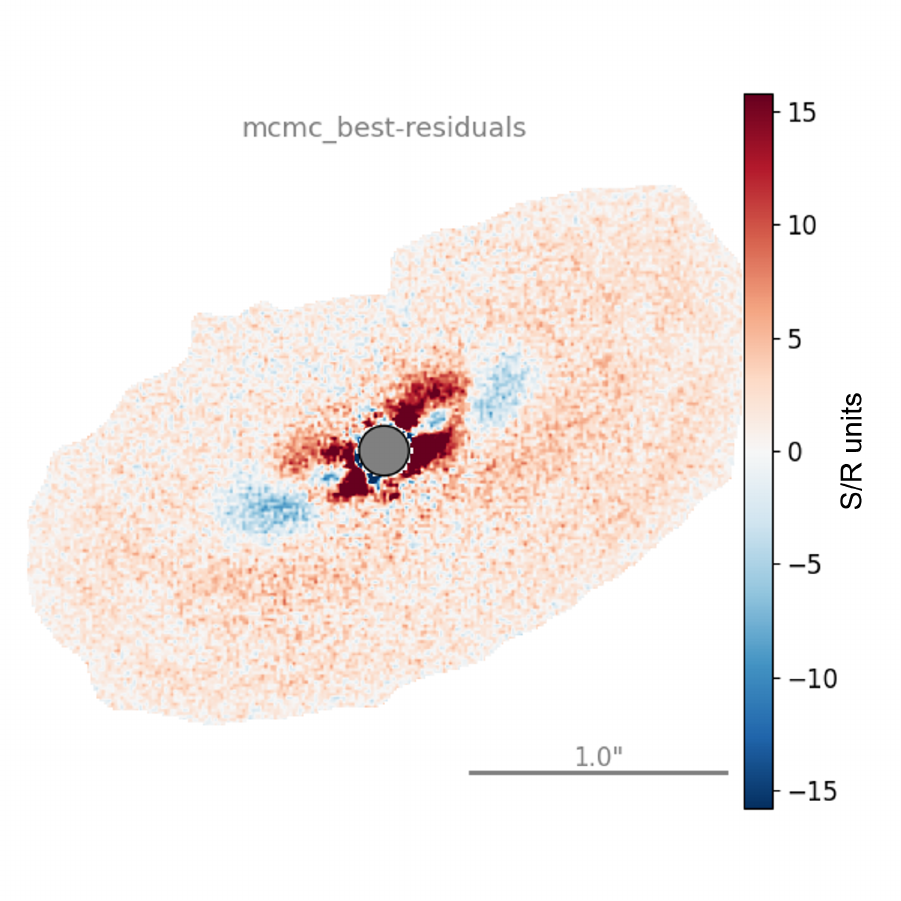}
                \caption{Best model residuals}
                \label{fig:BN_mcmc-res}
            \end{subfigure}
    
            \caption{Best model for BN\,CrA resulting from the MCMC procedure. Panel (a) illustrates the best model at $\lambda = \SI{1.625}{\um}$, convolved to the data resolution. The grey circle represents the coronagraph. Panel (b) shows the residuals between the data and the best model. North is up and east is left. }
            \label{fig:BN_mcmc}
        \end{figure}

        The disk around BN\,CrA is fit by a slightly flatter model than V721\,CrA (see \figurename~\ref{fig:BN_best-mcmc} and \tablename~\ref{tab:best-mcmc}) as we obtain a scale height $hr_0 \approx 0.07 $, although the flaring is slightly higher at $ \beta \sim 0.4 $. Most notably, its mass is two orders of magnitude lower than V721\,CrA. 
        For BN\,CrA  the best-fit dust mass is around \SI{1.0e-6}{\Msun}. It is possible that a significant mass fraction has already been trapped in larger solids than  we probed and modelled through $H$-band scattered light. This would be in line with the not-so-young age estimates for this disk, justifying the depletion of the micron-sized dust population. 
        In addition, since Class II disks have completely depleted their natal envelopes \citep[e.g.][]{Williams&Cieza2011:diskevo, Morbidelli+2024} and cannot be supplied with further gas or dust, the older age of BN\,CrA (versus V721\,CrA) can result in a decreased dust mass due to accretion and dispersal mechanisms. 
        
        The disk of BN\,CrA is compatible with an inclination of \ang{\sim 70}, which is close to (but more reliable than) the value obtained in the preliminary estimation in \secname~\ref{subsec:CrA_contourfitting}. Even this inclination value could be affected by the specific choice of dust opacity for the regression.  
        Notwithstanding the significant inclination, the midplane does not show a strong dark lane in the scattered light data. The parametric model also does not show a bottom side bright enough to be seen above the noise level (used as the minimum of the colour scale). 

        In the residuals image, illustrated in \figurename~\ref{fig:BN_mcmc-res}, we note two features. 
        First, there is evidence for a gap or shadow in the middle of the disk, which can be recovered even more clearly from the image in \figurename~\ref{fig:BN_r2}. This gap or shadow extends radially from approximately \SI{70}{au} to \SI{100}{au}. It is not possible to detect its far side (as for the entire disk), whereas the near side may be slightly contaminated by the brightness of the forward-scattering peak. 
        Second, our model settles in the middle between the bright inner region of the observed disk, the less bright outer part and the gap--shadow, struggling to closely follow the different parts of this structured disk. In this context, with the chosen dust opacity, the model is very bright along the major axis, generating a kind of x-shaped residual in the inner regions.

\section{Discussion}
    \label{sec:discuss}

    \begin{figure*}     
        \centering
        \begin{subfigure}[c]{0.49\textwidth}
            \centering
            \includegraphics[trim=0 0 0 0, clip, width=\textwidth]{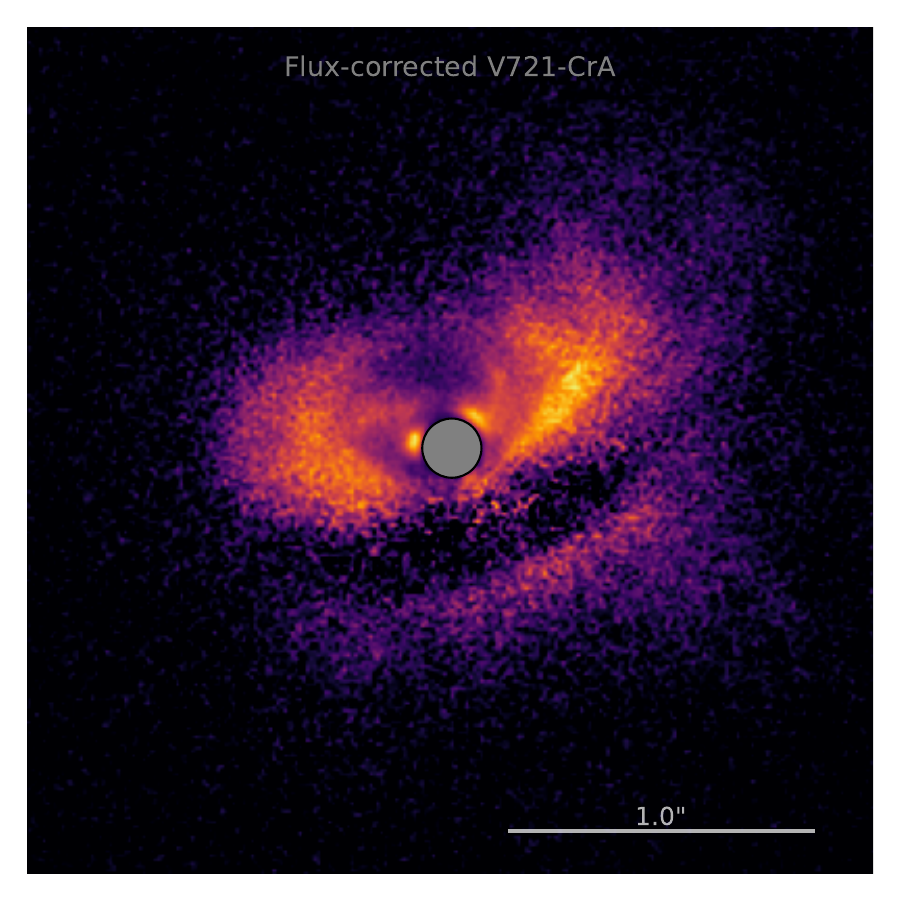}
            \caption{V721 CrA \Qphi{}$\cdot r^2$}
            \label{fig:V721_r2}
        \end{subfigure}
        \hfill
        \begin{subfigure}[c]{0.49\textwidth}
        \centering
            \includegraphics[trim=0 0 0 0, clip, width=\textwidth]{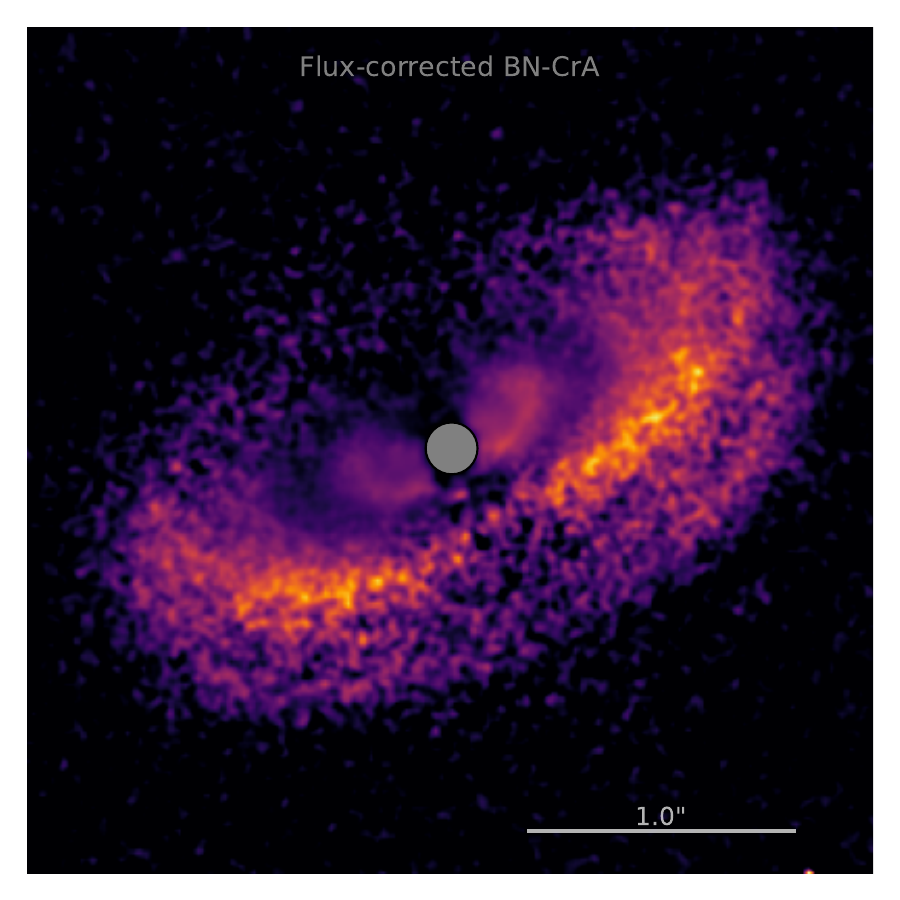}
            \caption{BN CrA \Qphi{}$ \cdot r^2 $}
            \label{fig:BN_r2}
        \end{subfigure}
        
        \caption{Flux-corrected \Qphi{} frames for V721\,CrA and BN\,CrA. 
        For panel (b) a light Gaussian smoothing was applied to the data to reduce the impact of the detector noise. 
        The colour scale is linear. The coronagraph area is covered by a grey circle. The rulers in the lower right corner of both panels indicates the scale of the FoV. North is up and east to the left. }
        \label{fig:CrA_r2-Qphis}
    \end{figure*}

    The parametric modelling described in \secname~\ref{sec:methods} allowed us to give a general morphological characterisation of V721\,CrA and BN\,CrA disks. 
    Adopting the best-fit parameters, we generated the  {flux-corrected} images, illustrated in \figurename~\ref{fig:CrA_r2-Qphis}, using \texttt{diskmap} \citep{Stolker+2016:diskmap}. In this context, the correction refers to the natural flux decay from the central star, going as $ F \propto 1 / r^2 $ (with $ r$ the distance to the light source). Multiplying the original data by a map of the $ r^2 $ of the whole disk surface, it is possible to more accurately assess the structural differences in the dust distribution. 
    These frames shed crucial insights into the details of the disk substructures. 
    
    V721\,CrA is the thicker and less axisymmetric of the two disks. Its inner regions are characterised by spiral arms and darker voids, in addition to two bright spots right on the edge of the coronagraphic mask. 
    The extent and shape of the spiral wakes are not trivial to assess, due to the relatively high inclination of V721\,CrA. It is not easy to understand whether the outer disk edge consists of spirals as well, also considering the puffed west disk side. 
    We note that the bright surface of this disk is better represented by $ 3 \times h_\mathrm{d}(r) $ (see \eqname~\ref{eq:h(r)}, the scale height of the models), which would be in agreement with \citet{Kenyon&Hartmann1987}, regarding the scattering surface of the small dust in an optically thick disk at NIR wavelengths around \SI{1}{\um}. 

    BN\,CrA is more axisymmetric, thinner, wider, and less bright than V721\,CrA. It has a darker lane approximately at half of its radius, which suggests the presence of ring-like substructures in this disk. The high inclination and the lower brightness of BN\,CrA imply that the far side is almost undetected (due to the weak polarised backscattering), whereas the front side displays a moderate forward-scattering peak.

    \subsection{Scattering phase functions}
        \label{subsec:SPF_discuss}

        \begin{figure}
            \centering      
            \includegraphics[width=\columnwidth]{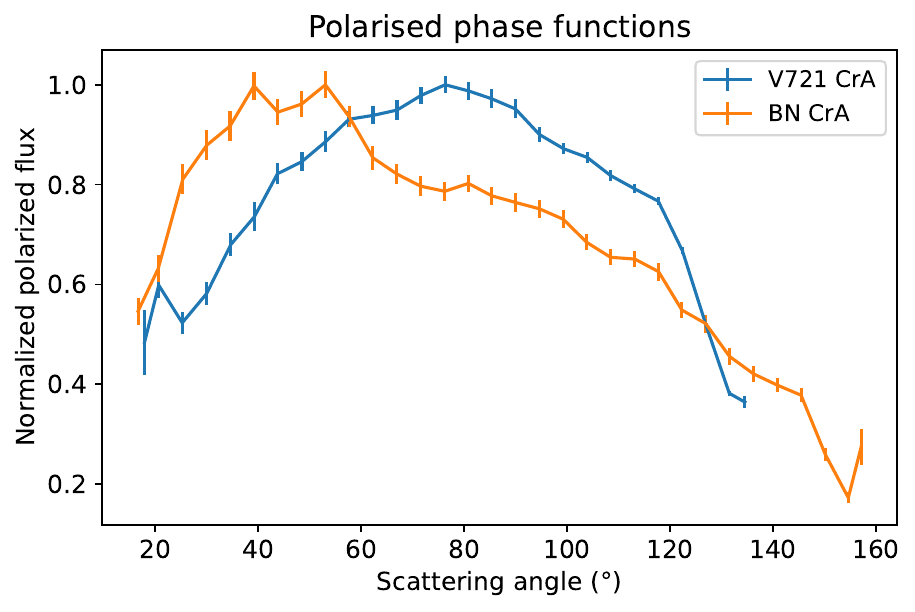}
    
            \caption{Polarised scattering phase functions for V721\,CrA and BN\,CrA, extracted from the flux-corrected \Qphi{}$ \cdot r^2$ frames. Each phase function was independently normalised. }
            \label{fig:phase_func_plot}
        \end{figure}

        The differences between the two disks are highlighted from the flux-corrected frames of \figurename~\ref{fig:CrA_r2-Qphis}.
        An important distinction, which was less obvious from the original data (see \figurename~\ref{fig:V721_Qphi} and \figurename~\ref{fig:BN_Qphi}), regards the shape of the scattering phase function (SPF). While BN\,CrA shows a prominent forward-scattering peak, V721\,CrA is definitely fainter on the front side. 
        To assess this difference quantitatively, we extracted the SPF of both disks, with \texttt{diskmap}. 
        Considering the high inclination of our disks, and the asymmetries in the detected substructures, it is not easy to obtain enough signal for the SPF in selected radial slices. Thus, we tried to use the maximum available disk surface, to extract the SPF averaged on the whole disk, 
        which means that we used the surface between $ \SI{40}{au} < r < \SI{100}{au} $ for V721\,CrA and between $ \SI{40}{au} < r < \SI{190}{au} $ for BN\,CrA. 
        The resulting SPF are plotted in \figurename~\ref{fig:phase_func_plot}. 
        The difference between the two disks is immediately evident in this plot. The forward scattering is  significantly stronger for BN\,CrA, whose SPF peaks around $ \theta \approx \ang{50} $. 
        Instead, V721\,CrA's phase function is more similar to a bell-shaped curve peaking very close to $ \theta \approx \ang{90} $. 

        The shape of the scattering phase function can be informative regarding the properties of the dust population (see e.g. \citealp{Min+2016:dust, Tazaki&Tanaka2018:dustscat, Tazaki&Dominik2022}). 
        A systematic comparison of the polarised SPF among ten circumstellar disks was done by \cite{Ginski+2023:SPF}. 
        If we compare their Fig.\,3 to our SPFs, we can see that most of their sample has an SPF qualitatively similar to that of BN\,CrA, with HD\,163296 and RXJ\,1615 being the most similar among all. 
        Interestingly, RXJ\,1615 is a faint disk with several rings and/or a dark lane in the middle, similar to BN\,CrA.
        The SPF of RXJ\,1615 (and others in their sample as well) also shows the backscattering peak that we recover for BN\,CrA. Assuming that these features are due to the properties of dust grains, it is likely that these disks share similar dust populations, which are compatible with low-porosity aggregates ($P_\mathrm{max} \approx 55\%$, \citealp{Ginski+2023:SPF}). 
        Moreover, the age estimate for HD\,163296 is around \SI{5}{Myr}, the same as for BN\,CrA (whereas the age of RXJ\,1615 is estimated at \SI{1.4}{Myr}, \citealp{Wahhaj+2010}). 
        
        On the other hand, the bell shape of V721\,CrA's SPF, with its weaker forward scattering, is generally associated with smaller and fluffier grains, with lower fractal dimension \citep[e.g][]{Min+2016:dust, Tazaki&Dominik2022}. This would be in line with the younger age estimated for V721\,CrA and consequently its lesser degree of dust evolution. 
        A caveat is that the SPF analysis described in this subsection implicitly assumes that the dust is homogeneously distributed in the azimuthal dimension. Any substructure or shadow would add uncertainty to the extraction of the SPF, as could be the case for both V721\,CrA and BN\,CrA. 
        
        In light of the differences between V721\,CrA and BN\,CrA that arise from the SPF analysis, future models could improve starting from the assumption of the same dust opacity for both disks (with compact grains). 
        The first step was mandatory since extracting the SPF requires assuming a disk geometry in the first place. 
        A further modelling iteration could shed more insights into the dust properties in each disk, but requires heavy computational resources, and is beyond the scope of the present work.

    \subsection{Surrounding environment}

       V721\,CrA and BN\,CrA belong to different subclusters within the Corona Australis complex, as we  discuss in the Introduction. The first one is on-cloud, whereas the second one is off-cloud. 
        We note here that the results of our analysis agree with the generally expected properties of disks in those groups. 

        On the one hand we have V721\,CrA, on-cloud, which shows the typical features of young disks \citep[see e.g.][]{Williams&Cieza2011:diskevo, Testi+2014:dustevo, Manara+2023:ppvii}. It is fit by thicker disk models (see \secname~\ref{subsec:V721_results}), and has a dust mass much higher than that of BN\,CrA. 
        Its dust  is possibly more abundant in smaller grains and/or monomers (see \secname~\ref{subsec:SPF_discuss}), as expected for its young age. 
        It shows signs of ongoing dynamical evolution with the spiral-like substructures identified from \figurename~\ref{fig:V721_r2}. 
        Interestingly, it has a large and faint asymmetrical feature on the north-west side. This might be a streamer of dust feeding the circumstellar disk (see e.g. T CrA \citealp{Rigliaco+2023:Tcra} or SU Aur \citealp{Ginski+2021:SUaur}) or otherwise a sparse outer spiral structure. 
        Both options are plausible, considering that the on-cloud group is  characterised by large reservoirs of interstellar material, available for feeding the newborn stars. 
        Overall, V721\,CrA seems more likely to interact with and be affected by its surrounding environment during its evolution. 
        In addition, it would be interesting to investigate the cause of the high RUWE\footnote{
        {Renormalised unit weight error}, it indicates the quality of the astrometric fit on the Gaia data.} of V721\,CrA ($\sim 2.1$). 
        A RUWE $ > 1.4$ is commonly interpreted as a possible hint of unresolved companions, indicating that the source is non-single or otherwise problematic.
        However, alternative explanations for a high RUWE, such as high stellar activity or circumstellar material \citep{Fitton+2022:ruwe} suggest a RUWE threshold of 2.5 for disk-bearing stars, as could be the case for V721\,CrA.
        Interferometric observations with GRAVITY, or spectroscopy for radial velocities, could be useful in searching for a potential stellar companion or ruling it out. 
        Lastly, if we consider the ALMA data, the millimetre-dust map obtained by \citet{Francis&vanDerMarel2020} shows a ring structure that is almost as extended as the dust observed in scattered light. On the other hand, the \element[13][]{CO} 2-1 map obtained by \citet{Woelfer+2023} shows gas emission at noticeably larger distances from the star. 
        Although gas is often traced well beyond dust, this indicates that the disk structure as a whole extends up to about 200\,au from the star (almost double the visible dust radius at about 120\,au). 
        
        On the other hand, we have BN\,CrA in the outskirts of the off-cloud group. 
        Its disk is thinner and less massive; it possibly consists of larger dust aggregates with lower porosity (see \secname~\ref{subsec:BN_results} and  \secname~\ref{subsec:SPF_discuss}), which suggests that its dust was processed more intensely. 
        The disk of BN\,CrA is more axisymmetric, and, contrary to V721\,CrA, it does not show signs of interaction with its surrounding environment. 
        The disk appears isolated in our observations, with no remarkable sparse emission above the noise in its vicinity. 
        Lastly, we did not detect any spiral structure in BN\,CrA.

\section{Summary}
    \label{sec:CrA_conclude}

    In this work, we presented new $H$-band polarised observations of two transition disks in the Corona Australis star-forming region: V721\,CrA and BN\,CrA. 
    These datasets resolved these disks for the first time in IR scattered light, showing two extended and structured disks. 
    We analysed the \Qphi{} frames to characterise their morphology and geometry, adopting simple parametric models and MCMC regressions. 
    Then, we used the best-fit parameters to generate the $r^2$-corrected maps of the dust, where the disk substructures can be best recovered, and extracted the scattering phase functions. 

    We summarise here the main results of our analysis: 
    \begin{itemize}
        \item V721\,CrA is a thick disk, with a radius of \SI{\sim 120}{au} and an estimated micron-sized dust mass around \SI{2.8e-4}{\Msun}. It shows spiral wakes most clearly at intermediate radii,  in agreement with the gas rotation maps. There is potential evidence for additional material infalling from the surroundings (to be further investigated), perhaps responsible for breaking the axial symmetry, but not the midplane symmetry. 
        \item BN\,CrA is thinner and is around \SI{200}{au} in size. It is more axisymmetric and shows a gap or shadow between \SIrange{70}{100}{au}. The micron-dust mass is estimated to be around \SI{1.0e-6}{\Msun}. 
        \item Their polarised scattering phase functions suggest that the two disks have slightly different dust populations, with V721\,CrA possibly having smaller grains and BN\,CrA having larger and more compact ones.  
        \item Both V721\,CrA and BN\,CrA have features that are in good agreement with the general properties expected, respectively,  for the age of the on- and off-cloud subgroups in CrA to which they belong. 
    \end{itemize}

    Considering the relatively high inclinations at which we observe these two disks (\ang{57.6} and \ang{70.6} for V721\,CrA and BN\,CrA, respectively), it might be difficult to detect  planets forming within them. The midplane, where protoplanets could be forming, is not easy to probe at these IR wavelengths because of all the dust in the disk atmosphere. 
    However, other techniques may be used to  study these objects further. Radial velocity monitoring, for example, could be better suited than imaging to detect potential substellar companions around these solar-mass stars. 
    
    For BN\,CrA, whose disk had not been previously resolved by any other observation, ALMA observations in the Band 6 or 3 continuum could provide insight into the larger solids and the midplane properties. This would allow  the comparison of the axisymmetrical features, seen in the IR-polarised light, with those observed in the millimetre-dust thermal continuum. As we recovered circular shadows and ring-like features, it seems plausible that the larger midplane solids could be structured as well. 
    
    Lastly, additional observations in scattered light, from the optical \Ha{} to the $K$-band, would be important to better constrain the scattering phase functions in both V721\,CrA and BN\,CrA, and thus their dust grain properties. 
    They would also provide additional data points to pinpoint the geometry and morphology of these two disks.

    \bigskip

    \vfill

\begin{acknowledgements}
    G.C. thanks N. Cuello for early comments and useful suggestions.
    This work has made use of data from the European Space Agency (ESA) mission Gaia (https://www.cosmos.esa.int/gaia), processed by the Gaia Data Processing and Analysis Consortium (DPAC, https://www.cosmos.esa.int/web/gaia/dpac/consortium). 
    This publication makes use of VOSA, developed under the Spanish Virtual Observatory (https://svo.cab.inta-csic.es) project funded by MCIN/AEI/10.13039/501100011033/ through grant PID2020-112949GB-I00.
    VOSA has been partially updated by using funding from the European Union's Horizon 2020 Research and Innovation Programme, under Grant Agreement nº 776403 (EXOPLANETS-A).
    E.R. acknowledges financial contribution from the PRIN-MUR 2022 20228JPA3A “The path to star and planet formation in the JWST era (PATH)” funded by NextGeneration EU", and from the INAF mini-grant RF 2022. 
    S.F. is funded by the European Union (ERC, UNVEIL, 101076613), and acknowledges financial contribution from PRIN-MUR 2022YP5ACE. Views and opinions expressed, however, are those of the author(s) only and do not necessarily reflect those of the European Union or the ERC. Neither the European Union nor the granting authority can be held responsible for them. 
    A.R. has been supported by the UK Science and Technology Facilities Council (STFC) via the consolidated grant ST/W000997/1 and by the European Union’s Horizon 2020 research and innovation programme under the Marie Sklodowska-Curie grant agreement No. 823823 (RISE DUSTBUSTERS project). 
    A.Z. acknowledges support from ANID -- Millennium Science Initiative Program -- Center Code NCN2024\_001 and Fondecyt Regular grant number 1250249.
    M.B. has received funding from the European Research Council (ERC) under the European Union’s Horizon 2020 research and innovation programme (PROTOPLANETS, grant agreement No. 101002188).
    
\end{acknowledgements}

\bibliographystyle{aa}
\bibliography{CrA}

\begin{appendix}

    \section{SED modelling}
        \label{subsec:sed_models}

\begin{figure}[b]
            \centering       
            \includegraphics[width=\columnwidth]{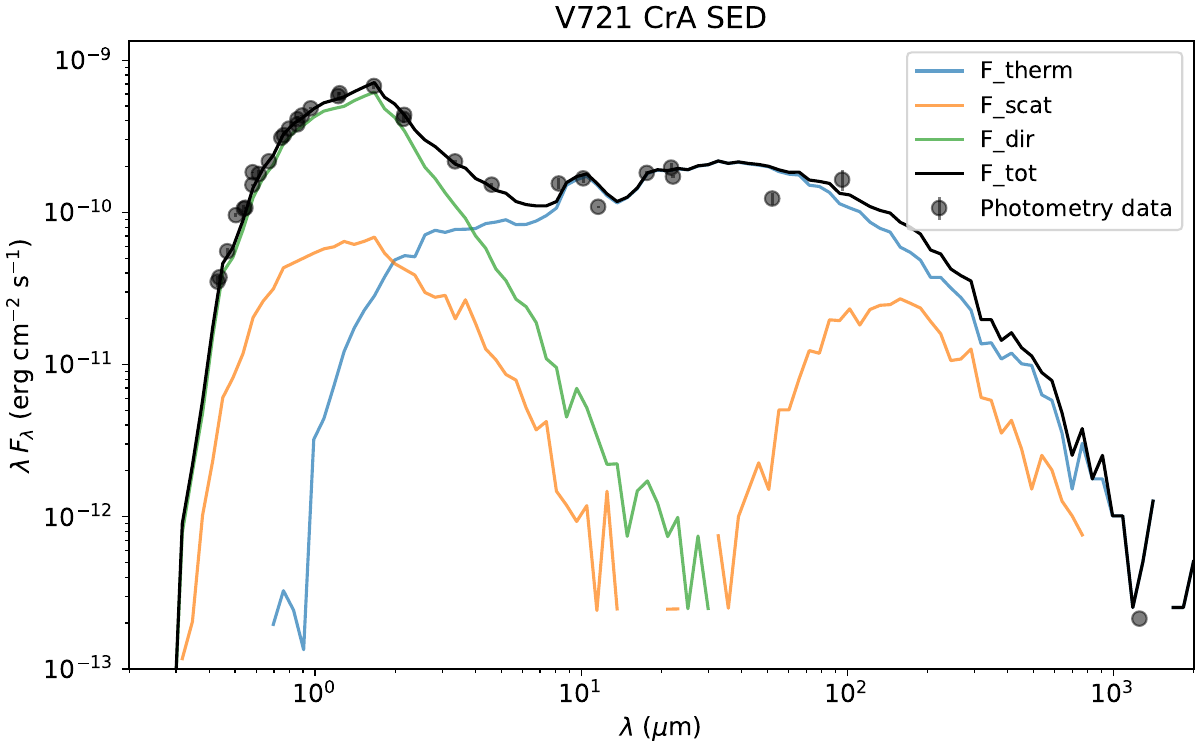}
    
            \caption{SED modelling for V721\,CrA. The different contributions to the total energy (thermal, scattering and direct stellar light) are plotted in separate colours, as indicated in the legend. The black line is the summed total energy, while the black circles show the photometry points obtained from literature. }
            \label{fig:V721_SED}
        \end{figure}

\begin{figure}[t]
            \centering       
            \includegraphics[width=\columnwidth]{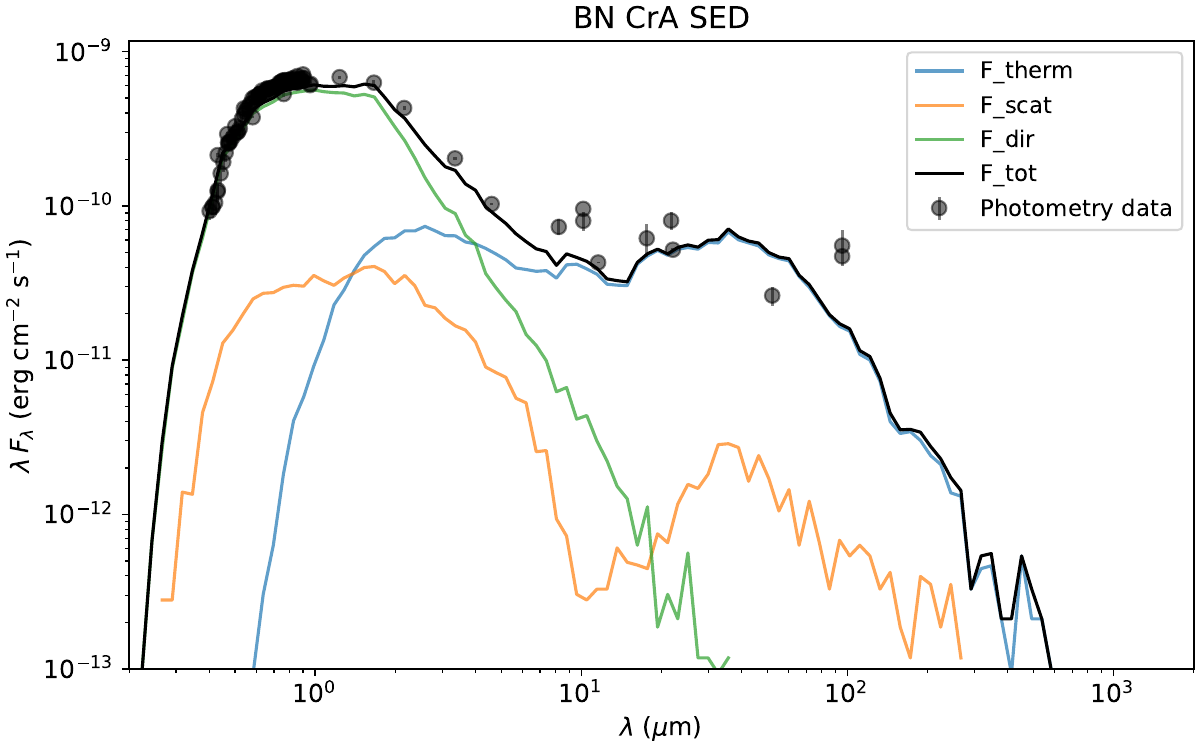}
    
            \caption{Same as Fig. A.1, but for BN\,CrA}
            \label{fig:BN_SED}
        \end{figure}

        Some hints on the gas and dust distribution around the two stars can be obtained by modelling their spectral energy distribution (SED), in particular regarding the dust and gas distribution in the region behind the coronagraph, where the direct imaging cannot provide information.
        To model the SED we use the dust radiative transfer model developed by \citet{Whitney+2003b, Whitney+2003a}. The code uses a Monte Carlo radiative transfer scheme that follows photon packages emitted by the central star as they are scattered, absorbed, and re-emitted throughout the disk.
        For the modelling process, we assume the stellar properties reported in \tablename~\ref{tab:star_data}. 
        For the disk mass and inclination we use the best-fit values reported in \tablename~\ref{tab:best-mcmc} for V721\,CrA and BN\,CrA, multiplying the dust mass by a factor of 100 to estimate the total disk mass. 
        The modelled SEDs are shown in \figurename~\ref{fig:V721_SED} and \figurename~\ref{fig:BN_SED}. The black points correspond to the observed photometric points, while the coloured curves show the different SED contributions that were computed, as labelled in the figure.
        In order to reproduce the observed SEDs, we assume the following disk characteristics. 
        The disk around V721\,CrA is slightly puffed, and the micron-sized dust grains imaged with SPHERE show that it extends from behind the coronagraph (\SI{1}{au}) up to \SI{\sim 120}{au}. 
        In the SED modelling we assume that small grains (grain size $\leq 200\AA$) are uniformly distributed along the disk. The large grains ($\geq 200 \AA$) are distributed from \SI{30}{au} up to \SI{120}{au}, with the inner edge chosen in order to be in agreement with the millimetre-dust gap found with ALMA by \cite{Francis&vanDerMarel2020}. 
        The disk around BN\,CrA extends from \SI{1}{au} to 200\,au, and shows a gap between 70\,au and 100\,au. No inner-disk gap is modelled in order to reproduce the observed SED. The disk is more evolved than the one around V721\,CrA, and this is in agreement with the older age of the star+disk system.
        In both cases, the SED modelling and the disk properties are consistent with a younger age for V721\,CrA, where the disk is puffier and it produces a stronger infrared excess, and an older age for BN\,CrA, with a smoother disk, less massive and with a potential gap in the middle.

    \FloatBarrier

    \section{Supplementary figures}

    We report more useful plots and figures in this appendix to avoid overcrowding the main text. Specifically, we include the figures regarding the MCMC sampling for the parametric models of both disks showing the post-burn-in steps of the walkers and the corner plots with the marginalised posterior distributions (\figurename~\ref{fig:V721_MCMCsampling} and \figurename~\ref{fig:BN_MCMCsampling}). 
    
    We also report the details of the prior distributions in our MCMC procedure (described in \secname~\ref{sec:methods}) in \tablename~\ref{tab:priors}. The priors were all flat distributions. The Range column shows the interval within which we would accept the new random sample.

    \begin{table}[h!]      
            \centering
            \caption{Priors ranges used for the MCMC exploration.  }
            \label{tab:priors}
            \renewcommand{\arraystretch}{1.2}
            \setlength{\tabcolsep}{14pt}
            \begin{tabular}{l | c}
                \toprule
                Parameter   & Range \\
                \midrule
                $ \log_{10}(M_\mathrm{d} / \si{\Msun})$  &  [-8, -2] \\ 
                $ hr_0 $   &    [0.01, 1.0]     \\ 
                $ \beta $  &    [0.01, 0.5]     \\ 
                $ i $ [°]  &    [20, 80]        \\ 
                \bottomrule
            \end{tabular}
            
    \end{table}

    \begin{figure*} 
        \centering     
        \begin{subfigure}[c]{0.7\linewidth}
            \centering
            \includegraphics[trim=0 0 0 0, clip, width=\textwidth]{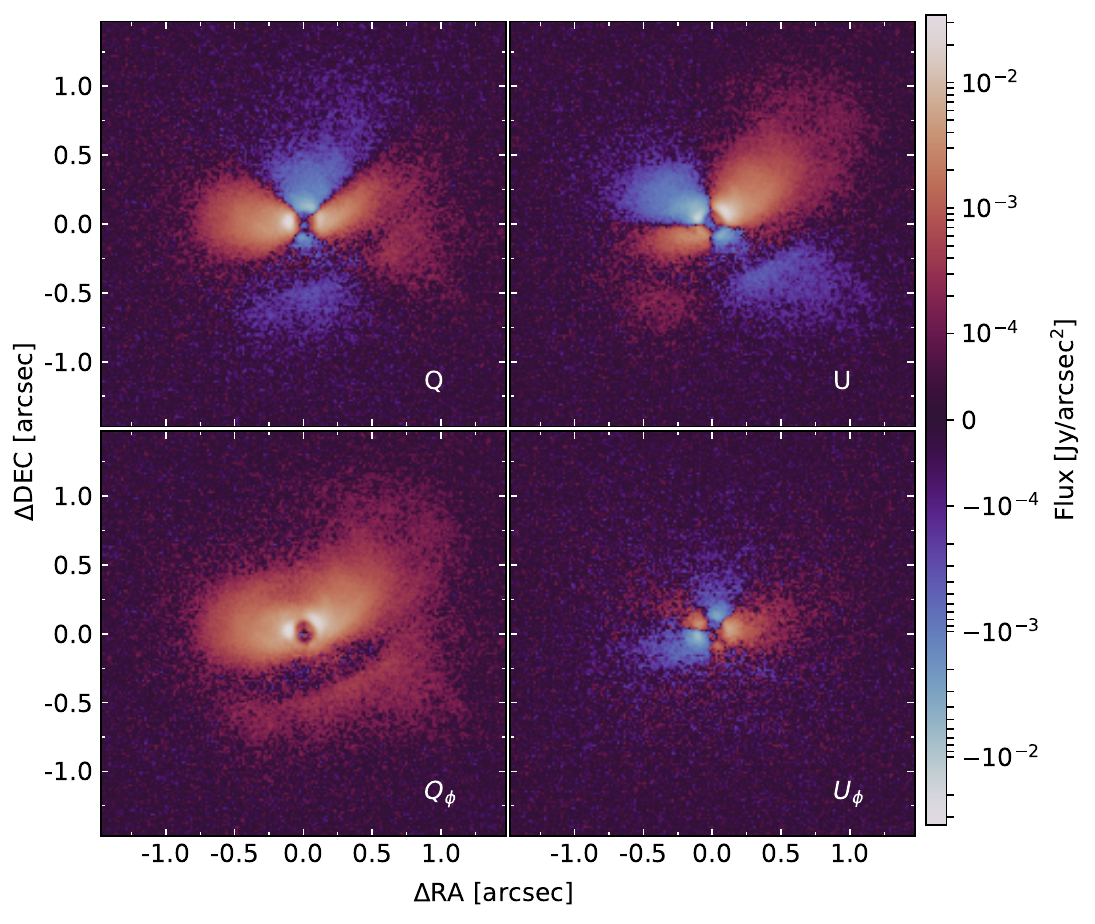}
            \caption{ \small V721 CrA }
            \label{fig:V721_quadStokes}
        \end{subfigure} 
        \bigskip
    
        \begin{subfigure}[c]{0.7\linewidth}
            \centering      
            \includegraphics[trim=0 0 0 0, clip, width=\textwidth]{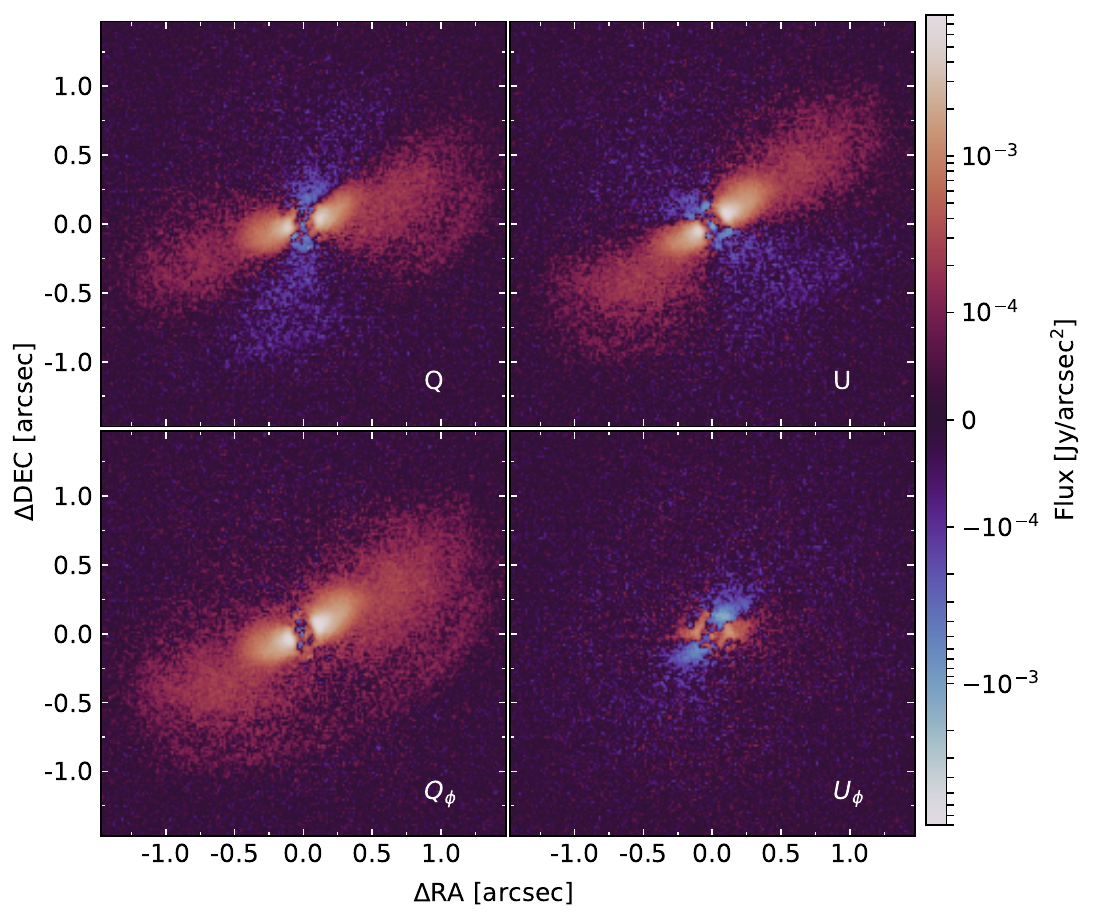}
            \caption{ \small BN CrA }
            \label{fig:BN_quadStokes}
        \end{subfigure}

        \caption{ IRDIS $H$-band polarimetric Stokes frames of V721\,CrA (a) and BN\,CrA (b). The frame name is given in the lower right corner of each panel. 
        The same colour scale is used in all the panels for each target and  is logarithmic. North is up and east to the left. }
        \label{fig:quadStokes}
    \end{figure*}

    \begin{figure} 
        \centering     
        \begin{subfigure}[c]{\columnwidth}
            \centering
            \includegraphics[trim=40 50 0 0, clip, width= \textwidth]{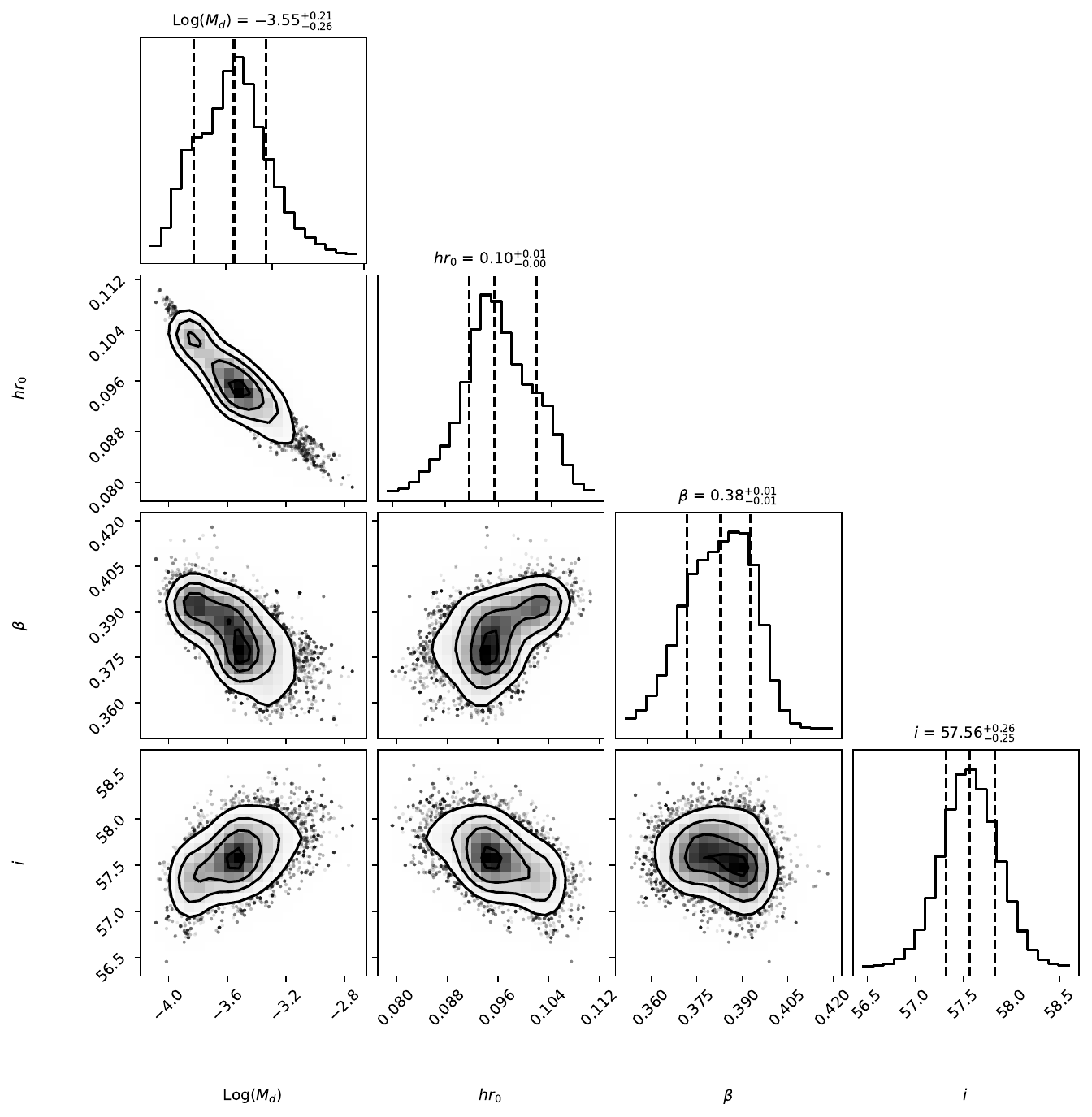}
            \caption{ Corner plot }
            \label{fig:V721_corner}
        \end{subfigure} 
        \bigskip
    
        \begin{subfigure}[c]{\columnwidth}
            \centering      
            \includegraphics[trim=0 0 0 0, clip, width= \textwidth]{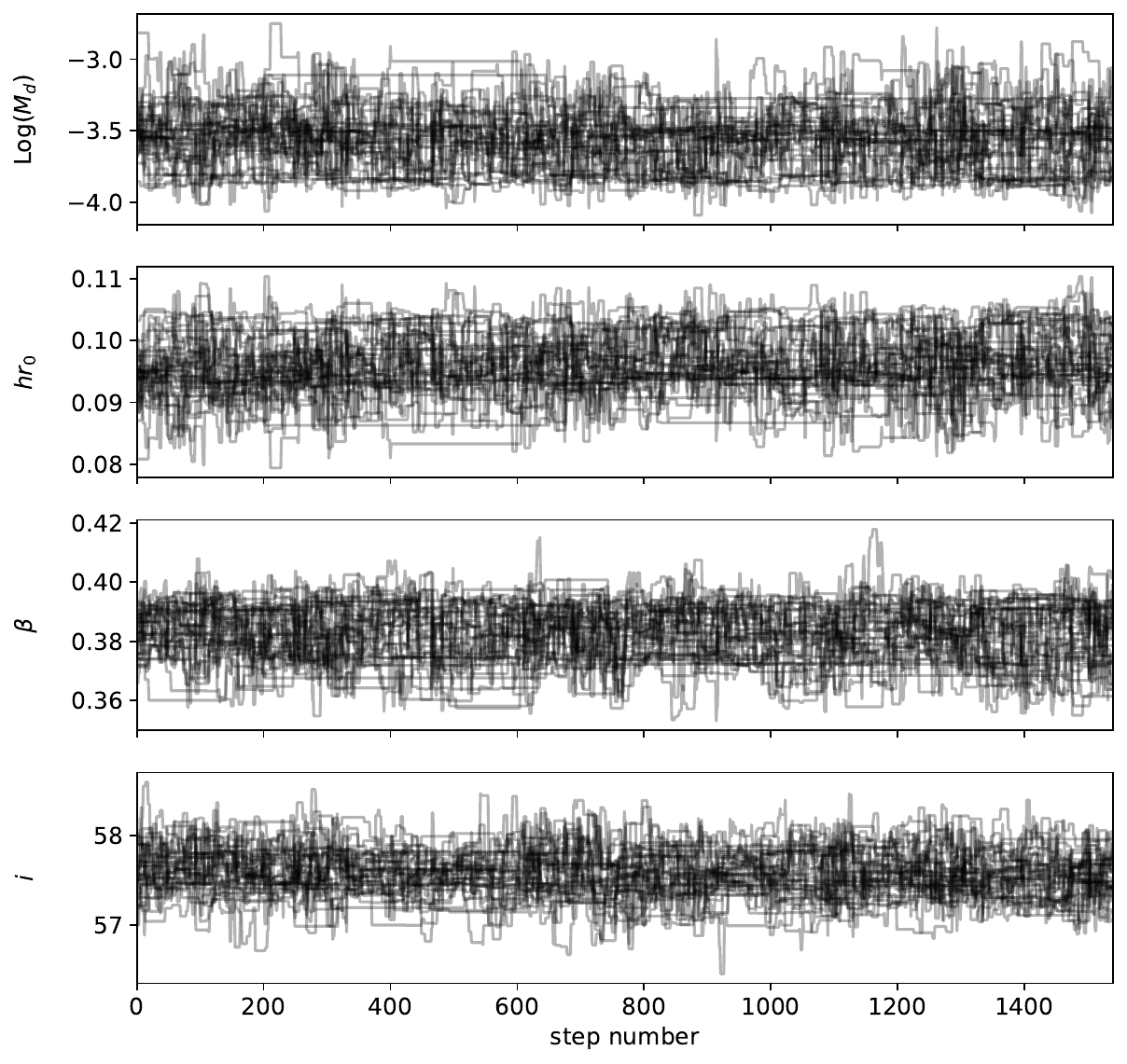}
            \caption{ Chains   }
            \label{fig:V721_chainsteps}
        \end{subfigure}

        \caption{Corner plot showing the posteriors and relative MCMC walkers chains (after removing the burn-in) for the V721\,CrA best model, illustrated in \figurename~\ref{fig:V721_mcmc}. }
        \label{fig:V721_MCMCsampling}
    \end{figure}

    \begin{figure}  
        \centering      
        \begin{subfigure}[c]{\columnwidth}
            \centering
            \includegraphics[trim=40 50 0 0, clip, width=\textwidth]{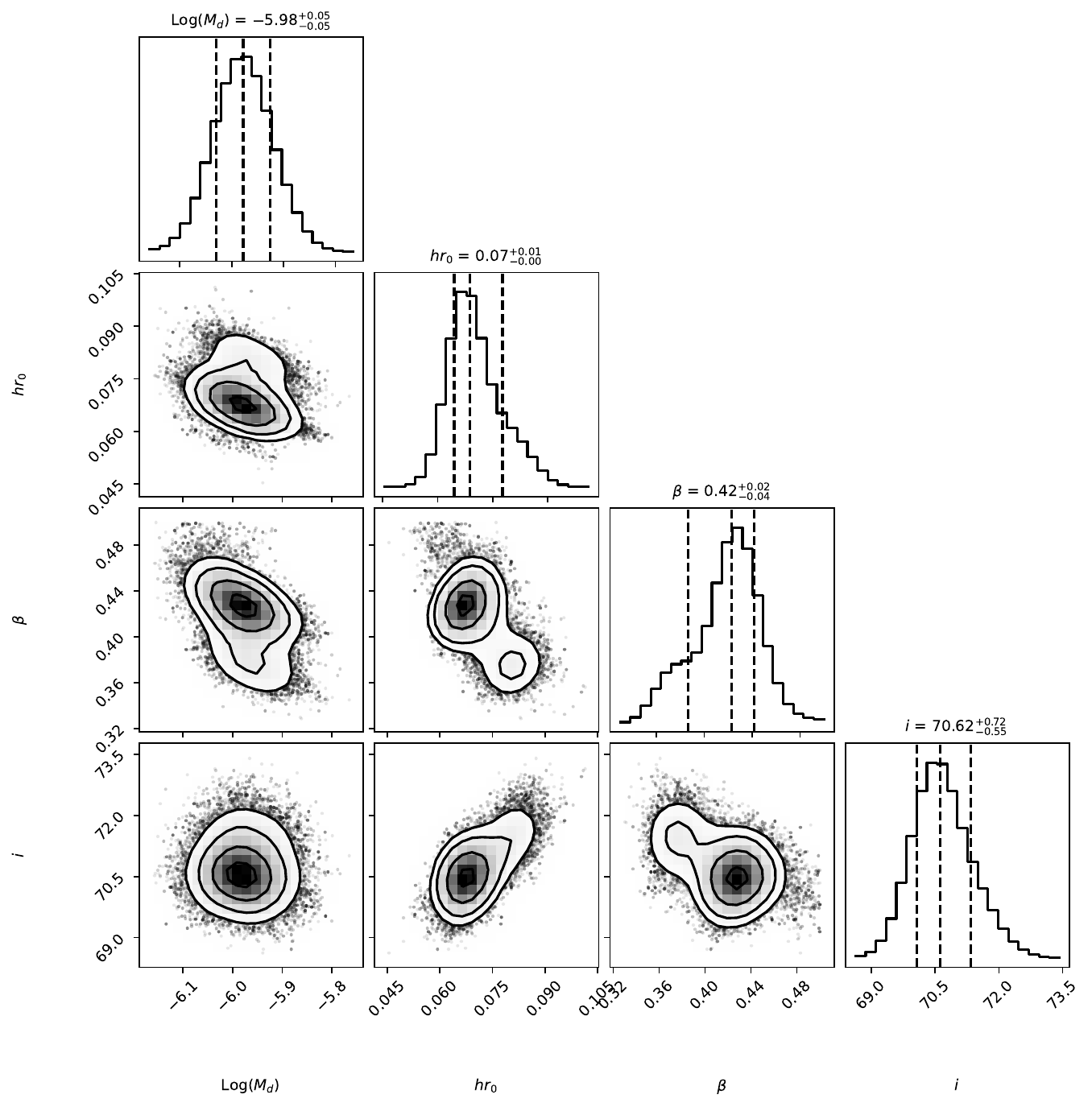}
            \caption{ Corner plot }
            \label{fig:BN_corner}
        \end{subfigure}
        \bigskip

        \begin{subfigure}[c]{\columnwidth}
            \centering      
            \includegraphics[trim=0 0 0 0, clip, width=\textwidth]{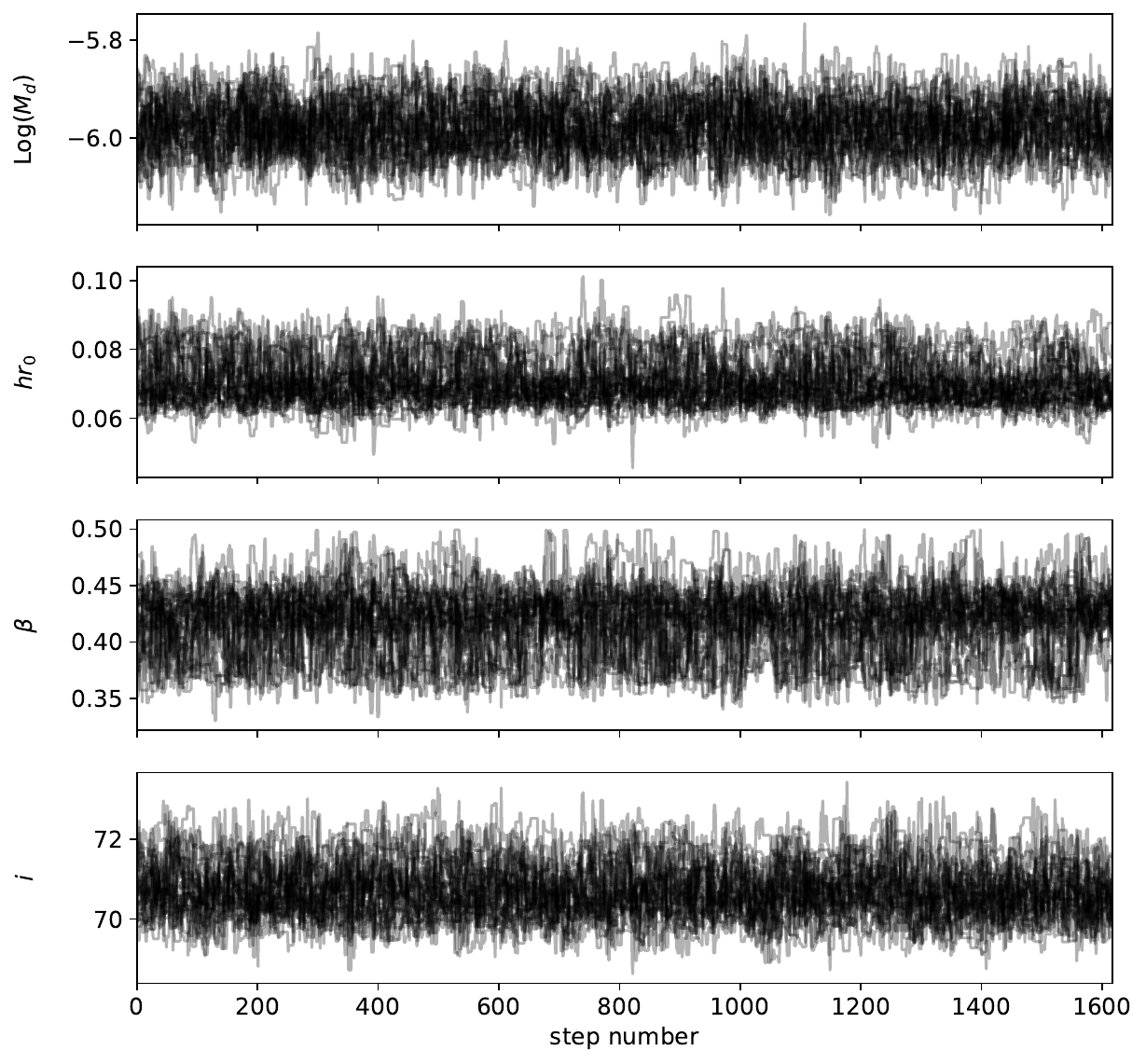}
            \caption{ Chains }
            \label{fig:BN_chainsteps}
        \end{subfigure}
        
        \caption{Corner plot showing the posteriors and relative MCMC walkers chains (after removing the burn-in) for the BN\,CrA best model, illustrated in \figurename~\ref{fig:BN_mcmc}. }
        \label{fig:BN_MCMCsampling}
    \end{figure}

\end{appendix}

\end{document}